\documentclass[final,5p,times,twocolumn]{elsarticle}

\makeatletter
\def\ps@pprintTitle{%
 \let\@oddhead\@empty
 \let\@evenhead\@empty
 \def\@oddfoot{\footnotesize\itshape
       Preprint accepted for publication in \ifx\@journal\@empty Elsevier
       \else\@journal\fi\hfill\today}%
 \let\@evenfoot\@oddfoot}
\makeatother

\usepackage[backgroundcolor=cyan]{todonotes}
\usepackage{caption}
\usepackage{subcaption}
\usepackage{mathtools}
\usepackage[]{nomencl}
\usepackage[nodots]{numcompress}
\usepackage{widetext}
\usepackage{import}
\usepackage[colorlinks,urlcolor=blue,bookmarks=false]{hyperref}

\def\xlinkspace#1 #2{%
 \ifx\relax#2%
 \xlinkdash#1-\relax
 \else
 \xlinkdash#1 -\relax
 \expandafter\xlinkspace\expandafter#2%
 \fi}

\def\xlinkdash#1-#2{%
 \ifx\relax#2%
 \tmp{#1}%
 \else
 \tmp{#1-}%
 \expandafter\xlinkdash\expandafter#2%
 \fi}

\makenomenclature

\usepackage{amssymb}
\usepackage{amsbsy}
\usepackage{amsmath}
\usepackage{siunitx}
\makeatletter
\def\ang{\kernel@ifnextchar[\ang@{\ang@[]}}
\def\ang@[#1]#2{\def\@tempb{[{#1}]}\@ang#2;;;\@nil}
\def\@ang#1;#2;#3;#4\@nil{%
  \@tempswatrue
  \@@ang{#1}{\degree}%
  \@@ang{#2}{\arcminute}%
  \@@ang{#3}{\arcsecond}%
}
\def\@@ang#1#2{%
  \edef\@tempa{#1}%
  \ifx\@tempa\@empty
   \if@tempswa
    \expandafter\SI\@tempb{0}{#2}%
   \fi
  \else
    \expandafter\SI\@tempb{#1}{#2}%
    \@tempswafalse
  \fi
}
\makeatother

\newcount\colveccount
\newcommand*\colvec[1]{
        \global\colveccount#1
        \begin{pmatrix}
        \colvecnext
}
\def\colvecnext#1{
        #1
        \global\advance\colveccount-1
        \ifnum\colveccount>0
                \\
                \expandafter\colvecnext
        \else
                \end{pmatrix}
        \fi
}

\newcount\colveccounts
\newcommand*\colvecs[1]{
        \global\colveccounts#1
        \begin{bmatrix}
        \colvecnexts
}
\def\colvecnexts#1{
        #1
        \global\advance\colveccounts-1
        \ifnum\colveccounts>0
                \\
                \expandafter\colvecnexts
        \else
                \end{bmatrix}
        \fi
}

\newcommand{\remark}[1]%
{
\textcolor{red}{#1}%
}%

\renewcommand*\arraystretch{1.1}
\let\oldhat\hat
\renewcommand{\vec}[1]{\mathbf{#1}}
\renewcommand{\hat}[1]{\oldhat{\mathbf{#1}}}

\newcommand{\admax}{\alpha_\mr{d,max}}
\newcommand{\ad}{\alpha_\mr{d}}
\newcommand{\as}{\alpha_\mr{s}}
\newcommand{\Aside}{A_\mr{side}}

\newcommand{\cs}{c_\mr{s}}
\newcommand{\ctoc}{c_\mr{2,c}}
\newcommand{\dt}{d_\mr{t}}
\newcommand{\dnr}{d_\mr{n,r}}

\newcommand{\hb}{h_\mr{b}}
\newcommand{\hk}{h_\mr{k}}
\newcommand{\id}{i_\mr{d}}
\newcommand{\is}{i_\mr{s}}
\newcommand{\isc}{i_\mr{s,c}}
\newcommand{\ksd}{K_\mr{s,D}}
\newcommand{\kds}{K_\mr{d,s}}
\newcommand{\ls}{l_\mr{s}}

\newcommand{\lti}{l_{\mr{t},i}}
\newcommand{\dotlti}{\dot{l}_{\mr{t},i}}
\newcommand{\mk}{m_\mr{k}}

\newcommand{\lso}{l_{\mr{s},0}}
\newcommand{\ti}{t_i}
\newcommand{\ud}{u_\mr{d}}
\newcommand{\udd}{u_{\mr{d},0}}
\newcommand{\udmax}{u_\mr{d,max}}
\newcommand{\us}{u_\mr{s}}
\newcommand{\uss}{u_{\mr{s},0}}
\newcommand{\vto}{v_\mr{t,o}}
\newcommand{\too}{v_\mr{o}}

\newcommand{\vw}{v_\mr{w}}
\newcommand{\vwg}{v_\mr{w,ref}}
\newcommand{\vwe}{v_\mr{w,exp}}
\newcommand{\wk}{w_\mr{k}}
\newcommand{\wrel}{w_\mr{rel}}
\newcommand{\zr}{z_\mr{ref}}

\newcommand{\vPKCU}{\vec{P}_\mr{KCU}}
\newcommand{\vaa}{\vec{a}}

\newcommand{\vfk}{\vec{F}_\mr{k}}
\newcommand{\vfs}{\vec{f}_\mr{s}}
\newcommand{\vfi}{\vec{f}_i}
\newcommand{\vfsi}{\vec{f}_{\mr{s},i}}
\newcommand{\vsi}{\vec{s}_i}
\newcommand{\vfsii}{\vec{f}_{\mr{s},i-1}}

\newcommand{\vdi}{\vec{d}_i}
\newcommand{\vdii}{\vec{d}_{i-1}}

\newcommand{\vD}{\vec{F}_\mr{D}}
\newcommand{\vG}{\vec{F}_\mr{g}}
\newcommand{\vL}{\vec{F}_\mr{L}}
\newcommand{\vp}{\vec{p}}
\newcommand{\vpi}{\vec{p}_i}
\newcommand{\vpii}{\vec{p}_{i+1}}

\newcommand{\vPc}{\vec{P}_\mr{c}}

\newcommand{\vrp}{\vec{R}_\mr{p}}
\newcommand{\vrv}{\vec{R}_\mr{v}}
\newcommand{\vsnn}{\vec{s}_{n-1}}

\newcommand{\vsvi}{\vec{s}_{\mr{v},i}}

\newcommand{\vS}{\vec{F}_\mr{s}}
\newcommand{\vvi}{\vec{v}_i}
\newcommand{\vvasi}{\vec{v}_{\mr{a,s},i}}
\newcommand{\vvasip}{\vec{v}_{\mr{a,s},i,\bot}}
\newcommand{\vvaxy}{\vec{v}_{\mr{a},xy}}
\newcommand{\vvaxz}{\vec{v}_{\mr{a},xz}}

\newcommand{\vvii}{\vec{v}_{i+1}}

\newcommand{\vvsi}{\vec{v}_{\mr{s},i}}
\newcommand{\vvwsi}{\vec{v}_{\mr{w,s},i}}
\newcommand{\vvwl}{v_\mr{w,log}}
\newcommand{\vvwk}{\vec{v}_\mr{w,k}}

\newcommand{\mr}[1]{{\mathrm{#1}}}

\newcommand{\mA}{m_\mr{A}}
\newcommand{\mB}{m_\mr{B}}
\newcommand{\mC}{m_\mr{C}}
\newcommand{\mD}{m_\mr{D}}
\newcommand{\mKCU}{m_\mr{KCU}}

\newcommand{\vFk}{{\vec{F}_\mr{k}}}

\newcommand{\vFLB}{{\vec{F}^\mr{B}_\mr{L}}}
\newcommand{\vFDB}{{\vec{F}^\mr{B}_\mr{D}}}
\newcommand{\vFLC}{{\vec{F}^\mr{C}_\mr{L}}}
\newcommand{\vFDC}{{\vec{F}^\mr{C}_\mr{D}}}
\newcommand{\vFLD}{{\vec{F}^\mr{D}_\mr{L}}}
\newcommand{\vFDD}{{\vec{F}^\mr{D}_\mr{D}}}

\newcommand{\vex}{{\vec{e}_x}}
\newcommand{\vey}{{\vec{e}_y}}
\newcommand{\vez}{{\vec{e}_z}}

\newcommand{\vexx}{{\vec{e}_{x,0}}}
\newcommand{\veyy}{{\vec{e}_{y,0}}}
\newcommand{\vezz}{{\vec{e}_{z,0}}}

\newcommand{\CD}{{C_\mr{D}}}
\newcommand{\CL}{{C_\mr{L}}}

\newcommand{\vvk}{{\vec{v}_\mr{k}}}

\newcommand{\vvw}{{\vec{v}_\mr{w}}}

\newcommand{\vva}{{\vec{v}_\mr{a}}}
\newcommand{\va}{{v_\mr{a}}}
\newcommand{\vasquared}{{v^2_\mr{a}}}

\newcommand{\vrk}{\vec{r}_{\rm{k}}}

\newcommand{\zk}{z_\mr{k}}

\newcommand{\xw}{{x_\mr{w}}}
\newcommand{\yw}{{y_\mr{w}}}
\newcommand{\zw}{{z_\mr{w}}}

\newcommand{\lt}{{l_\mr{t}}}

\newcommand{\alphad}{\alpha_\mr{d}}
\newcommand{\alphazero}{\alpha_0}
\newcommand{\alphas}{\alpha_\mr{s}}
\newcommand{\alphaszero}{\alpha_{\mr{s},0}}
\newcommand{\alphasmax}{\alpha_{\mr{s, max}}}
\newcommand{\alphadmax}{\alpha_{\mr{d,max}}}
\newcommand{\ato}{a_{\mr{t,o}}}
\newcommand{\alphaB}{\alpha_\mr{B}}
\newcommand{\alphaC}{\alpha_\mr{C}}
\newcommand{\alphaD}{\alpha_\mr{D}}
\newcommand{\vaCxy}{v_{\mr{a,C,}xy}}
\newcommand{\vaBxz}{v_{\mr{a,B,}xz}}
\newcommand{\vaDxy}{v_{\mr{a,D,}xy}}
\newcommand{\vaB}{v_{\mr{a,B}}}
\newcommand{\vaC}{v_{\mr{a,C}}}
\newcommand{\vaD}{v_{\mr{a,D}}}
\newcommand{\ye}{\vec{Y}_\mr{e}}
\newcommand{\yezero}{\vec{Y}_\mr{e,0}}
\newcommand{\dotye}{\dot{\vec{Y}}_\mr{e}}
\newcommand{\dotyezero}{\dot{\vec{Y}}_\mr{e,0}}
\newcommand{\rre}{\vec{R}_\mr{e}}
\newcommand{\rrr}{R_\mr{r}}
\newcommand{\re}{\vec{r}_\mr{e}}
\newcommand{\dotvto}{\dot{v}_\mr{t,o}}
\newcommand{\taud}{\tau_\mr{d}}
\newcommand{\taug}{\tau_\mr{g}}
\newcommand{\tauf}{\tau_\mr{f}}
\newcommand{\taus}{\tau_\mr{s}}
\newcommand{\taum}{\tau_\mr{m}}
\newcommand{\cf}{c_\mr{f}}
\newcommand{\vsn}{v_\mr{s,n}}
\newcommand{\vs}{v_\mr{s}}
\newcommand{\en}{E_\mr{n}}
\newcommand{\cdt}{c_\mr{d,t}}
\newcommand{\umax}{u_\mr{max}}
\newcommand{\sigmaf}{\sigma_\mr{f}}
\newcommand{\etacyc}{\eta_\mr{cyc}}
\newcommand{\pav}{p_\mr{av}}
\newcommand{\etap}{\eta_\mr{p}}
\newcommand{\fto}{F_\mr{t,o}}
\newcommand{\fti}{F_\mr{t,i}}
\newcommand{\vti}{v_\mr{t,i}}

\journal{Renewable Energy}

\begin{document}

\begin{frontmatter}


\title{Dynamic Model of a Pumping Kite Power System}
\author{Uwe Fechner\corref{cor1}}
\ead{u.fechner@tudelft.nl}
\author{Rolf van der Vlugt}
\author{Edwin Schreuder}
\author{Roland Schmehl}
\address{Delft University of Technology, Faculty  of Aerospace Engineering, Kluyverweg 1, 2629HS Delft, Netherlands}
\cortext[cor1]{Corresponding author. Tel.: +31 15 278 8902.}

\begin{abstract}
Converting the traction power of kites into electricity can be a low cost solution for wind energy.
Reliable control of both trajectory and tether reeling is crucial. 
The present study proposes a modelling framework describing the dynamic behaviour of the interconnected system components, suitable for design and optimization of the control systems.
The wing, bridle, airborne control unit and tether are represented as a particle system using spring-damper elements to describe their mechanical properties. 
Two kite models are proposed: a point mass model and a four point model.
Reeling of the tether is modelled by varying the lengths of constituent tether elements.  Dynamic behaviour of the ground station is included. The framework is validated by combining it with the automatic control system used for the operation of a kite power system demonstrator. The simulation results show that the point mass model can be adjusted to match the measured behaviour during a pumping cycle.
The four point model can better predict the influence of gravity and inertia on the steering response and remains stable also at low tether forces.  
Compared to simple one point models, the proposed framework is more accurate and robust while allowing real-time simulations of the complete system.
\end{abstract}


\end{frontmatter}


\begin{thenomenclature} 

 \nomgroup{Symbols}

  \item [{$c$}]\begingroup damping coefficient of tether segment [Ns/m]\nomeqref {0}
		\nompageref{1}
  \item [{$c_0$}]\begingroup unit damping coefficient [Ns]\nomeqref {0}
		\nompageref{1}
  \item [{$\cs$}]\begingroup steering coefficient (one point kite model) [-]\nomeqref {0}
		\nompageref{1}
  \item [{$\dt$}]\begingroup tether diameter [m]\nomeqref {0}
		\nompageref{1}
  \item [{$\id$}]\begingroup relative depower input of kite control unit (0, 1) [-]\nomeqref {0}
		\nompageref{1}
  \item [{$\is$}]\begingroup relative steering input of kite control unit (-1, 1) [-]\nomeqref {0}
		\nompageref{1}
  \item [{$k$}]\begingroup spring constant of tether segment [N/m]\nomeqref {0}
		\nompageref{1}
  \item [{$k_0$}]\begingroup unit spring constant [N]\nomeqref {0}
		\nompageref{1}
  \item [{$\ksd$}]\begingroup steering-induced drag coefficient [-]\nomeqref {0}
		\nompageref{1}
  \item [{$\lti$}]\begingroup tether length at beginning of time step $i$ [m]\nomeqref {0}
		\nompageref{1}
  \item [{$\mKCU$}]\begingroup mass of kite control unit [kg]\nomeqref {0}
		\nompageref{1}
  \item [{$\mk$}]\begingroup mass of kite [kg]\nomeqref {0}
		\nompageref{1}
  \item [{$n$}]\begingroup number of tether segments [-]\nomeqref {0}
		\nompageref{1}
  \item [{$\lso$}]\begingroup initial length of tether segment [m]\nomeqref {0}
		\nompageref{1}
  \item [{$\ud$}]\begingroup relative depower setting of kite control unit (0, 1) [-]\nomeqref {0}
		\nompageref{1}
  \item [{$\us$}]\begingroup relative steering setting of kite control unit (-1, 1) [-]\nomeqref {0}
		\nompageref{1}
  \item [{$\too$}]\begingroup tether reel-out speed [m/s]\nomeqref {0}
		\nompageref{1}
  \item [{$\vwg$}]\begingroup horizontal wind velocity at 6~m height [m/s]\nomeqref {0}
		\nompageref{1}
  \item [{$z$}]\begingroup height of kite or tether segment [m]\nomeqref {0}
		\nompageref{1}
  \item [{$\vaa$}]\begingroup vector of accelerations of tether particles [m/s$^2$]\nomeqref {0}
		\nompageref{1}
  \item [{$\vdi$}]\begingroup drag force vector of tether segment $i$\nomeqref {0}
		\nompageref{1}
  \item [{$\vG,\vS$}]\begingroup vectors of the gravity and steering forces of kite [N]\nomeqref {0}
		\nompageref{1}
  \item [{$\vL,\vD$}]\begingroup lift and drag force vectors of kite [N]\nomeqref {0}
		\nompageref{1}
  \item [{$\vp$}]\begingroup vector of positions of tether particles [m]\nomeqref {0}
		\nompageref{1}
  \item [{$\vec{A},\vec{B}$}]\begingroup position vectors of the front and top kite particles  [m]\nomeqref {0}
		\nompageref{1}
  \item [{$\vec{C},\vec{D}$}]\begingroup position vectors of the right and left kite particles [m]\nomeqref {0}
		\nompageref{1}
  \item [{$\vec{R}$}]\begingroup vector of the residual of the implicit problem/ model\nomeqref {0}
		\nompageref{1}
  \item [{$\vsi$}]\begingroup vector from the tether particle i to the particle i+1 [m]\nomeqref {0}
		\nompageref{1}
  \item [{$\vsvi$}]\begingroup velocity of tether particle $i+1$ relative to particle $i$ [m/s]\nomeqref {0}
		\nompageref{1}
  \item [{$\vva$}]\begingroup vector of apparent air velocity [m/s]\nomeqref {0}
		\nompageref{1}
  \item [{$\vvwk$}]\begingroup vector of wind velocity at the height of kite [m/s]\nomeqref {0}
		\nompageref{1}
  \item [{$\vex,\vey,\vez$}]\begingroup unit vector of the x, y and z-axis of the kite-reference frame\nomeqref {0}
		\nompageref{1}
  \item [{$\boldsymbol{Y}$}]\begingroup state vector of the implicit problem/ model\nomeqref {0}
		\nompageref{1}
  \item [{$\alpha, \beta$}]\begingroup angle of attack and elevation angle [rad]\nomeqref {0}
		\nompageref{1}
  \item [{$\rho$}]\begingroup air density [kg$m^{-3}$]\nomeqref {0}
		\nompageref{1}

\end{thenomenclature}
\section{Introduction}
\label{sec:introduction}
\noindent Wind energy is a major source of renewable energy. However, conventional wind turbines are restricted by physical and economic limits. Airborne wind energy has the potential to overcome some of the limitations, using tethered flying devices to reach altitudes of 400 to 600 m where the wind is stronger and steadier \cite{Archer2014}. The fact that airborne wind energy systems do not require towers reduces the costs per installation significantly.

The focus of this paper is the modelling of airborne wind energy systems that use the traction power of a tethered inflatable wing in a pumping cycle, as described in \cite{Vlugt2013} and \cite{Fechner2013}. The main components of such a single-tether kite power system (KPS) are the wing, the kite control unit (KCU) suspended below the wing by means of a bridle system, the tether and the drum-generator module, which is part of the ground station.
It is the objective to develop a system model that is real-time capable and of sufficient accuracy for the development and verification of flight path and ground station controllers. 

A dynamic model of a two-line kite is derived in \cite{Diehl2001}. Variations of the angle of attack are not taken into account and the simplicity of the model allows for an analytical derivation of a state space representation based on four dynamic states.
Further expanding on this model, \cite{Ahmed2011} proposed a kite power system model with three degrees of freedom (DOF), in which the kite is represented as a point mass at the end of the straight tether of variable length. Assuming a rigid wing with constant aerodynamic properties, the steering forces are derived as functions of the roll angle.

A discretisation of the tether as a multibody system has been proposed by \cite{Williams2007e}, using a Lagrangian approach to derive the equations of motion in generalised coordinates.
The advantage of this approach is the direct incorporation of constrains which results in a compact problem formulation.
This model used rigid tether segments, connected by spherical joints, which is not sufficient for modelling the tether force and implementing the force control loop. In addition it is adding and removing point masses during the simulation to simulate reel-out and reel-in of the tether. According to our experience this causes artificial discontinuities in the model which makes it difficult to implement the force control loop. For the kite it also used a point mass model. 

A model that uses a discretised tether with point masses connected by springs was published in \cite{Gohl2013}. The aerodynamics of the kite were modelled using the vortex lattice method, which means that it is using an advanced kite model. On the other hand it was not mentioned if the dynamics of the winch were modelled at all and no details were published on the question how reeling in and out was modelled. Other authors presented detailed generator and winch models \cite{Ahmed2011,Coleman2013}, but no or only a very simple model for the kite and the tether.

Coupling fluid and structural dynamic solvers for wind turbine applications has been studied by \cite{Vire2012,Vire2012a}, while fluid-structure interactions methods have been applied to kite aeroelastic behaviour by \cite{Bosch2014}. These kind of models might be useful for the design of improved kites, but they are very computational intensive and currently at least one order of magnitude slower than real-time~\cite{Bosch2014}. 

This paper presents a model where the dynamics of all major system components - the tether, the kite and the generator - are taken into account, with a focus on a novel discretised tether model which allows smooth reel-in and reel-out. It is soft real-time capable and thus suitable for the training of kite pilots and winch operators, but can also be used for software in the loop testing of KPS control systems, the development of estimation algorithms and for the optimization of flight trajectories. 

An improved one-point kite model is presented, that allows to change the angle of attack during simulation time and uses look-up tables to calculate the lift and drag as function of the angle of attack. It also takes the increased drag when flying around corners into account. In addition it uses a correction term to match the influence of gravity. This model can already be sufficient for optimizing flight trajectories.

For controller development a four-point kite model is devised, the most simple point mass model that has rotational inertial in all axis. This avoids discontinuities in the kite orientation which make the one-point kite model uncontrollable in curtain flight manoeuvres. In addition it is very close to a fully physical model: Many model parameters like the height and width of the kite and the height of the bridle can just be measured and do not need to be identified. Only the steering sensitivity parameters need to be identified because they depend on the flexibility of the kite which is not explicitly modelled.

This article will first explain the atmospheric model, then the tether model and the two kite models and finally the winch model. Furthermore, the control system is briefly explained. Subsequently a systematic approach for the model calibration is presented, with the goal to match the conditions of a real flight as good as possible. 

In the results section major parameters like force, speed, power and flight trajectory as obtained from the point mass model and the four point model are compared with data, measured using the Hydra kite of Delft University of Technology. Finally conclusions are drawn about the performance and accuracy of the described models and which improvements are still needed.

\section{Computational approach}
\noindent One of the requirements when building the model was, that it has to be (soft-) real-time capable. On the other hand, the programming effort should be limited and it should be easy to adapt the model to different kite power systems. It was found that high-level modelling tools like Simulink or Modellica were not capable to simulate a discretised tether that is reeling in or out in real-time. Therefore general purpose programming language was used that makes low-level optimizations of the modelling code possible. 

We are modelling the kite and the tether as a particle system, using discrete point masses which are connected by spring-damper elements. This has the advantage of a coherent model structure for which efficient mathematical methods for solving the stiff equation system exist \cite{Eberhardt2000}. For describing the positions of the particles a ground fixed reference frame is used, where the x-axis is pointing east, the y-axis north and the z-axis upwards. The origin is placed at the ground station.

The state vector of the system was constructed using the states of the tether particles, the states of the kite particles (only needed for the the four point kite model, because otherwise the last tether particle also represents the kite) and the scalar states of the winch (generator). Because no accurate, real-time measurements of the wind speed at the height of the kite were available, an atmospheric model, describing the wind profile, was also needed.

\subsection{Atmospheric model}
\noindent To determine the wind speed $\vw$ at the height of the kite and at the height of each tether segment, the power law \cite{Stull2000} and the log law \cite[p. 19]{Burton2001} are used. Input parameters are the ground wind speed $\vwg$ and the current height $z$ of the kite or tether segment. The ground wind speed used in this paper was measured at $\zr~=~6.0~\mr{m}$.  The power law establishes the relationship between $\vw$ and $\vwg$ as
\begin{equation}
\vwe = \vwg ~ \left(\frac{z}{\zr}\right)^\alpha
\label{eq:powerLaw}
\end{equation}
with the exponent $\alpha$ as fitting parameter.
The logarithmic law, which according to \cite[p. 20]{Burton2001} is more accurate than the least-square power law, can be written in the following form
\begin{equation}
\vvwl = \vwg ~ \frac{\log(z / z_0)}{\log(\zr / z_0)} ,
\label{eq:logLaw}
\end{equation}
where $\zr$ is the reference height and $z_0$ is the roughness length.
For this paper not only the ground wind speed $\vwg$ is measured, but once per flight additionally the wind speed at two more heights, $z_1$ and $z_2$. Then, a wind profile is fitted to these three wind speeds. To make a fit with three (speed, height) pairs possible, Eqns. (\ref{eq:logLaw}) and (\ref{eq:powerLaw}) are combined in the following way
\begin{equation}
\vw = \vvwl + K~(\vvwl - \vwe) .
\label{eq:windProfile}
\end{equation}
The fit is done by varying the surface roughness $z_0$ and $K$ until $v_w$ according to Eq.~(\ref{eq:windProfile}) matches the measured wind speed at all three heights. The exponent $\alpha$ is chosen according to
\begin{equation}
\alpha = \frac{\log(~\vwe(z_1)~/~\vwg~)}{\log(z) - \log(\zr)} ,
\label{eq:alpha}
\end{equation}
which results in $~\vwe(z_1)~=~ \vvwl(z_1)$.

\begin{figure}[ht]
\centering
\includegraphics[width=0.8\linewidth]{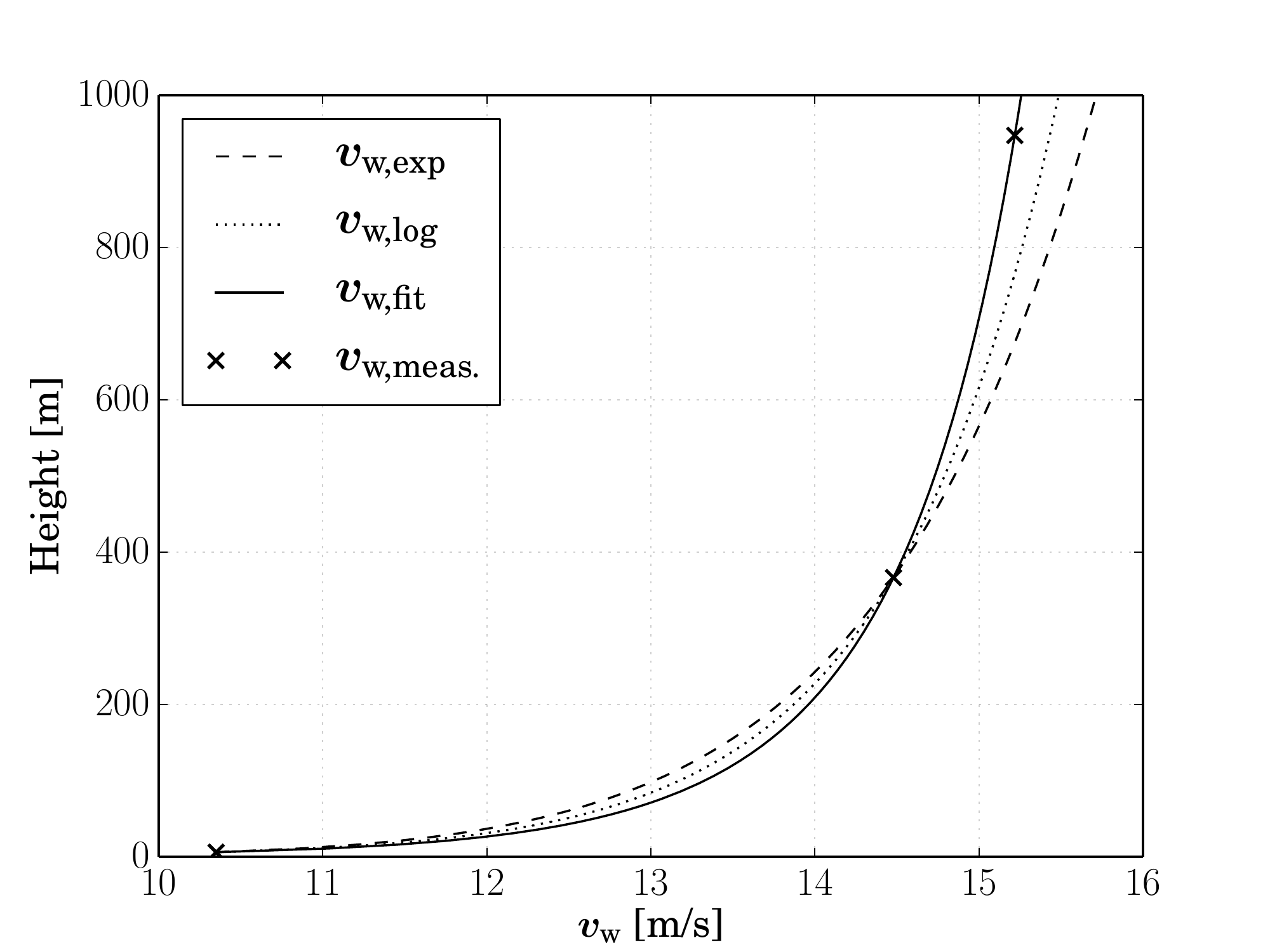}
\caption{Wind profile according the logarithmic law (dotted), the power law (dashed) and the fitted wind profile (solid), a linear combination of the others. Cross symbols represent measured values.}
\label{fig:wind_profile}
\end{figure}

An average sea-level density of $\rho_0 = 1.225$ kg/m$^3$ is assumed, and the height dependency is calculated according to
\begin{equation}
\rho = \rho_0 ~ \exp\left(\frac{z}{H_{\rho}}\right) ,
\label{eq:rho}
\end{equation}
where $z$ is the height and $H_{\rho} = 8.55$ km. An example for a fitted wind profile is shown in Fig. \ref{fig:wind_profile}, using the parameters from Table~\ref{tab:simulation_parameters}.

\subsection{Tether model}
\label{sec:tether_model}

The tether is modelled as a fixed number of lumped masses, connected with n spring damper elements as shown in Fig. \ref{fig:FourPointKite}. To simulate reel-in and reel-out the initial length of the tether segments $\ls$ is varied according to
\begin{equation}
\ls = \frac{\lti}{n} + \frac{\vto ~ (t-\ti)}{n} ,
\label{eq:inital_length}
\end{equation}
where $\lti$ is the tether length at the beginning of the i-th time step, $\vto$ the reel-out velocity, $t$ the simulation time and $\ti$ the simulation time at the beginning of the i-th time step.

This length is then used to calculate the spring and damping constants
\begin{equation}
k = {k_0}~\frac{l_0}{\ls} ,
\label{eq:spring_constant}
\end{equation}
\begin{equation}
c = {c_0}~\frac{l_0}{\ls} ,
\label{eq:damping_constant}
\end{equation}
where $l_0$ is the initial length of the tether segments at the beginning of the simulation.
The differential equations of the particle system are formulated as an implicit problem
\begin{align}
F(t,\vec{Y},\dot{\vec{Y}})~&=~0 , \label{eq:main_equation}\\
\vec{Y}(t_0)~&=~\vec{Y}_0 , \\
\dot{\vec{Y}}(t_0)~&=~\dot{\vec{Y}}_0 .
\end{align}
The state vector $\vec{Y}$ of the particle system is defined as 
\begin{equation}
\vec{Y} ~=~ (\vec{p}, \vec{v}) ,
\end{equation}
where $\vec{p}$ and $\vec{v}$ comprise the positions and velocities of the particles, respectively. For solving the problem only the residual $\vec{R}=F(t,\vec{Y},\dot{\vec{Y}})$ needs to be programmed.
The vector $\vec{R}$ consists of two partitions, the residual of the position vectors and its derivatives, and the residual of the velocity vectors and its derivatives, 
\begin{equation}
\vec{R}~=~(\vrp,~\vrv) .
\end{equation} 
The first partition can be calculated from Newton's law, $\vec{a}~=~\dot{\vec{v}}$. To calculate the second partition the particle forces must be known. On each particle are acting the forces of the spring damper element above and below the particle. In addition, half of the aerodynamic drag forces of the tether segments above and below of each particle have to be taken into account.

With $\vfsi$ denoting the tensile force of segment $i$ and $\vdi$ denoting the aerodynamic drag force of this segment as calculated in Eq.~(\ref{eq:drag_force}), the forces acting on the i-th particle can be calculated according to
\begin{equation}
\vfi~=~{\vfsii+{\vfsi}~+~\frac{1}{2}~(\vdi+\vdii)}~.
\label{eq:force}
\end{equation}
The first and the last particle have to be treated differently: For $i=0$ the lower spring force has to be replaced with the tether force as experienced by the ground station, and for the last tether particle the aerodynamic force of the kite has to be taken into account.

The spring forces are calculated according to Hooke's law
%
\begin{equation}
\vfs~=~{\left(k ~(\parallel \vsi \parallel -~\ls)~+~c~\left(\frac{\vsi}{\parallel \vsi \parallel} \cdot \vsvi \right)\right) \frac{\vsi}{\parallel \vsi \parallel}~~~} ,
\end{equation}
%
with $\ls$, $k$ and $c$ calculated according to Eqns. (\ref{eq:inital_length}), (\ref{eq:spring_constant}), (\ref{eq:damping_constant}) and with
\begin{align}
\vsi  & = \vpii - \vpi~, \\
\vsvi & = \vvii - \vvi~.
\end{align}
We use linear springs with a different stiffness for the extension and compression regimes. The stiffness for compression is much lower to model the behaviour of flexible bridle and tether lines, yet provide some structural stability.

The aerodynamic drag of any tether segment is calculated in the following way: First the wind speed at the height of the i-th tether segment $\vvwsi$ is calculated using Eq.~(\ref{eq:windProfile}).
Then, the average segment velocity is calculated as
\begin{equation}
\vvsi = \frac{1}{2}~(\vvii + \vvi) ,
\end{equation}
which leads to the apparent air velocity
\begin{equation}
\vvasi = \vvwsi - \vvsi .
\end{equation}
The drag of a cylinder is mainly caused by the component of $\vvasi$ that is perpendicular to the tether segment $\vec{s}_\mr{i}$ calculated as
\begin{equation}
\vvasip~=~\vvasi~-~\left(\vvasi \cdot \frac{\vsi}{\parallel \vsi \parallel}\right)~\frac{\vsi}{\parallel \vsi \parallel}~.
\end{equation}
Using this the drag force on the tether segment is resulting in
\begin{equation}
\vdi~=~\frac{1}{2} ~ c_\mr{d,t}~\rho ~ \parallel \vvasip \parallel ~ \parallel \vsi \parallel~d_\mr{t} ~{\vvasip},
\label{eq:drag_force}
\end{equation}
where $c_\mr{d,t}$  is the tether drag coefficient and $d_\mr{t}$ the tether diameter.

\subsection{Point mass kite model}
\label{subsec:point-mass_model}
\noindent The point mass model proposed in \cite[pp. 139--144]{Diehl2001} represents the kite as a discrete mass moving under the action of an aerodynamic force vector. It is also denoted as ``one point" or ``1p" model.
Steering is incorporated by an aerodynamic side force which depends linearly on the steering input.
This model does not account for rotational inertia, assuming that the wing is always aligned with the local relative flow experienced during flight.
Expanding on the original work, the model presented in the following allows for tether deformation and a variable angle of attack.

\subsubsection*{Reference frame}
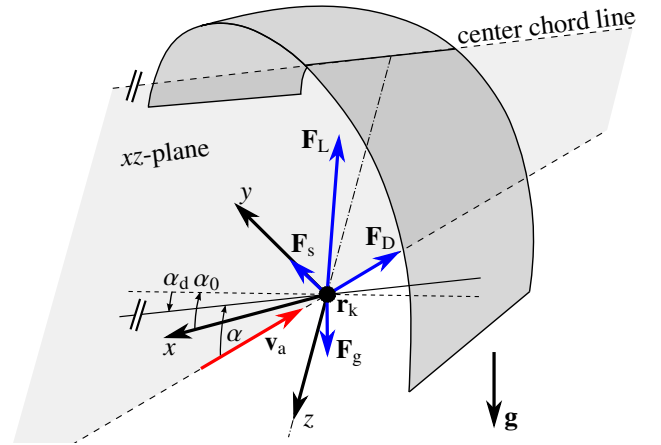
\begin{figure}[ht]
\centering
\def\svgwidth{0.45\textwidth}
\begingroup%
  \makeatletter%
  \providecommand\color[2][]{%
    \errmessage{(Inkscape) Color is used for the text in Inkscape, but the package 'color.sty' is not loaded}%
    \renewcommand\color[2][]{}%
  }%
  \providecommand\transparent[1]{%
    \errmessage{(Inkscape) Transparency is used (non-zero) for the text in Inkscape, but the package 'transparent.sty' is not loaded}%
    \renewcommand\transparent[1]{}%
  }%
  \providecommand\rotatebox[2]{#2}%
  \ifx\svgwidth\undefined%
    \setlength{\unitlength}{464bp}%
    \ifx\svgscale\undefined%
      \relax%
    \else%
      \setlength{\unitlength}{\unitlength * \real{\svgscale}}%
    \fi%
  \else%
    \setlength{\unitlength}{\svgwidth}%
  \fi%
  \global\let\svgwidth\undefined%
  \global\let\svgscale\undefined%
  \makeatother%
  \begin{picture}(1,0.70571463)%
    \put(0,0){\includegraphics[width=\unitlength]{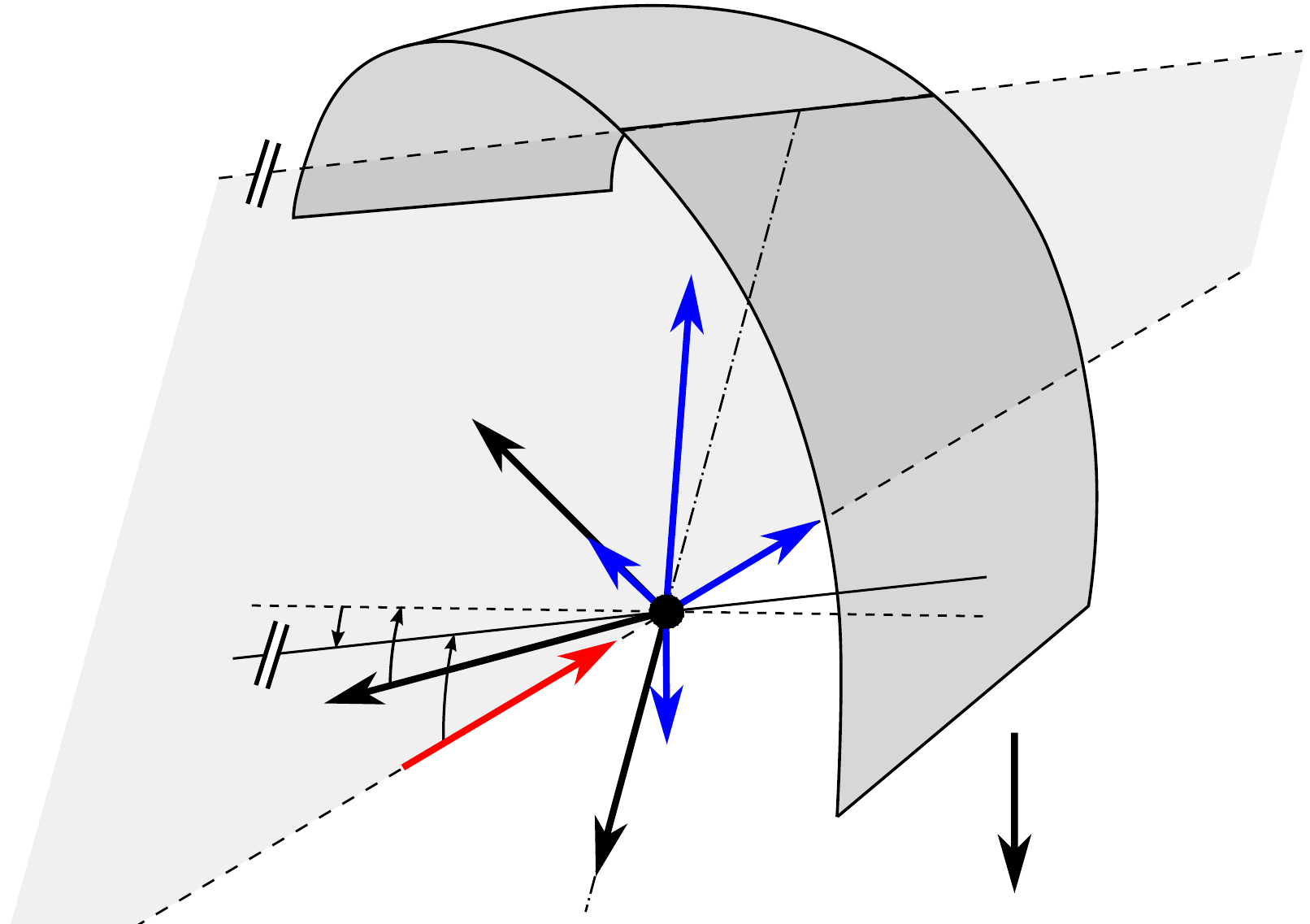}}%
    \put(0.52309783,0.13566052){\color[rgb]{0,0,0}\makebox(0,0)[lb]{\smash{$\vG$}}}%
    \put(0.52397632,0.21066957){\color[rgb]{0,0,0}\makebox(0,0)[lb]{\smash{$\vrk$}}}%
    \put(0.40889866,0.14865481){\color[rgb]{0,0,0}\makebox(0,0)[lb]{\smash{$\vva$}}}%
    \put(0.47125528,0.02732639){\color[rgb]{0,0,0}\makebox(0,0)[lb]{\smash{$z$}}}%
    \put(0.37025204,0.39141717){\color[rgb]{0,0,0}\makebox(0,0)[lb]{\smash{$y$}}}%
    \put(0.29652592,0.25200616){\color[rgb]{0,0,0}\makebox(0,0)[lb]{\smash{$\alphazero$}}}%
    \put(0.24819902,0.25390188){\color[rgb]{0,0,0}\makebox(0,0)[lb]{\smash{$\alphad$}}}%
    \put(0.34901797,0.16299502){\color[rgb]{0,0,0}\makebox(0,0)[lb]{\smash{$\alpha$}}}%
    \put(0.24657469,0.14116669){\color[rgb]{0,0,0}\makebox(0,0)[lb]{\smash{$x$}}}%
    \put(0.1793676,0.44498375){\color[rgb]{0,0,0}\rotatebox{6.73182406}{\makebox(0,0)[lb]{\smash{$xz$-plane}}}}%
    \put(0.46805146,0.47298171){\color[rgb]{0,0,0}\makebox(0,0)[lb]{\smash{$\vL$}}}%
    \put(0.56999222,0.31736892){\color[rgb]{0,0,0}\makebox(0,0)[lb]{\smash{$\vD$}}}%
    \put(0.45157671,0.30673491){\color[rgb]{0,0,0}\makebox(0,0)[lb]{\smash{$\vS$}}}%
    \put(0.79163634,0.02851584){\color[rgb]{0,0,0}\makebox(0,0)[lb]{\smash{$\vec{g}$}}}%
    \put(0.71812262,0.64055205){\color[rgb]{0,0,0}\rotatebox{6.73182406}{\makebox(0,0)[lb]{\smash{center chord line}}}}%
  \end{picture}%
\endgroup%
\caption[Kite reference frame]{Kite reference frame ($x,y,z$) of the point mass kite model. The physical wing is included here for the purpose of illustrating the concept of angle of attack and the assumed alignment with the relative flow.}
\label{fig:KiteReferenceFrame}
\end{figure}

\noindent
The kite reference frame ($x,y,z$) is defined on the basis of the local tether geometry and relative flow conditions.
As illustrated in Fig. \ref{fig:KiteReferenceFrame} the $z$-axis is aligned with the last tether segment. 
The $x$- and $y$-axes are constructed such that the apparent air velocity vector $\vva = \vvw - \vvk$ is in the $xz$-plane. 
This is based on the assumption that the wing is always aligned with the apparent wind velocity and that the sideslip velocity vanishes correspondingly.
The vector base is calculated as
\begin{align}
\vez & =~- ~\frac{\vsnn} {\parallel \vsnn \parallel} , \label{eq:z_vector} \\
\vey & =~\frac{\vva \times \vez}{\parallel \vva \times \vez \parallel} , \label{eq:y_vector} \\
\vex & =~\vey \times \vez . \label{eq:x_vector}
\end{align}
%
The unit vector $\vex$ is also called \emph{heading}, because it describes the orientation of the wing.

\subsubsection*{External forces}
\noindent The external force $\vFk$ acting on the point mass representation of the kite comprises contributions of aerodynamic lift $\vL$ and drag $\vD$, the aerodynamic side force $\vS$ and the gravitational force $\vG$
\begin{align}
\vL &~=~ \frac{1}{2} ~\rho~ \vasquared A~\CL(\alpha)~ \frac{\vva \times \vey}{\parallel \vva \times \vey \parallel} , \\
\vD &~=~ \frac{1}{2}~\rho~\vasquared A~\CD(\alpha) ~ (1 + \ksd~|\us|)  ~\frac{\vva}{\parallel\vva\parallel} , \\
\vS &~=~ \frac{1}{2}~\rho~\vasquared A~ \frac{\Aside}{A}~\cs~(\is + \isc)~\vey , \\
\vG &~=~ (\mk + \mKCU) ~ \vec{g} , \\
\vfk &~=~\vL+\vD+\vS+\vG . \label{eq:kite_force}
\end{align}

\noindent It should be emphasised that the drag force increases as the kite is steered due to kite deformation. Also, the steering force is based on the side area of the kite rather than the top area of the kite. The factored term ${\Aside}/{A}$ represents a parametrized description of a kite's geometry. The constant $\cs$ describes the steering sensitivity of the kite and has to be determined experimentally. The influence of the steering on the drag is described by $\ksd$. The empirical value of $\ksd~=~0.6$ is used. 
The variable $\isc$ is a correction term for the influence of gravity on the turn rate of the kite. It is calculated as follows
\begin{equation}
\isc~=~\frac{\ctoc}{\va}~\sin \psi~\cos~\beta .
\label{eq:correction}
\end{equation}
Equation (\ref{eq:correction}) is derived from the turn rate law as presented in Eq. (\ref{eq:psi}). The correction factor $\ctoc$ must be chosen such that the identified parameter $c_2$ of the one-point model matches the measurements. Without this correction the influence of gravity in this model was more than a factor of two higher.

\subsubsection*{Calculation of lift and drag as function of the angle of attack}
\begin{figure}[ht]
\centering
{\small
\def\svgwidth{0.35\textwidth}
\begingroup%
  \makeatletter%
  \providecommand\color[2][]{%
    \errmessage{(Inkscape) Color is used for the text in Inkscape, but the package 'color.sty' is not loaded}%
    \renewcommand\color[2][]{}%
  }%
  \providecommand\transparent[1]{%
    \errmessage{(Inkscape) Transparency is used (non-zero) for the text in Inkscape, but the package 'transparent.sty' is not loaded}%
    \renewcommand\transparent[1]{}%
  }%
  \providecommand\rotatebox[2]{#2}%
  \ifx\svgwidth\undefined%
    \setlength{\unitlength}{345.19174805bp}%
    \ifx\svgscale\undefined%
      \relax%
    \else%
      \setlength{\unitlength}{\unitlength * \real{\svgscale}}%
    \fi%
  \else%
    \setlength{\unitlength}{\svgwidth}%
  \fi%
  \global\let\svgwidth\undefined%
  \global\let\svgscale\undefined%
  \makeatother%
  \begin{picture}(1,1.0360725)%
    \put(0,0){\includegraphics[width=\unitlength]{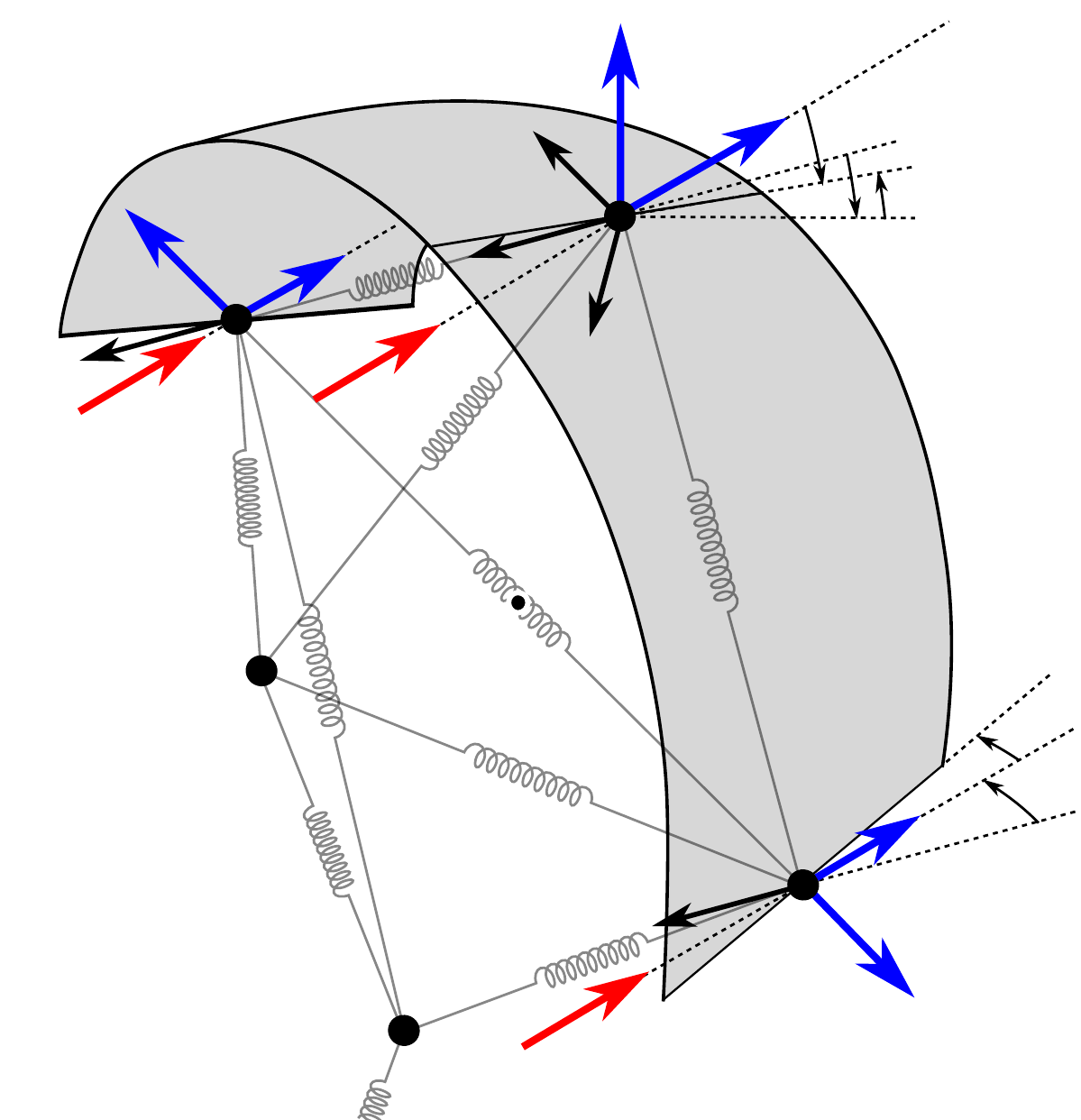}}%
    \put(0.6805526,0.94376364){\color[rgb]{0,0,0}\makebox(0,0)[lb]{\smash{$\vFDB$}}}%
    \put(0.58950044,0.9834522){\color[rgb]{0,0,0}\makebox(0,0)[lb]{\smash{$\vFLB$}}}%
    \put(0.45430066,0.88826772){\color[rgb]{0,0,0}\makebox(0,0)[lb]{\smash{$y$}}}%
    \put(0.55691686,0.71683516){\color[rgb]{0,0,0}\makebox(0,0)[lb]{\smash{$z$}}}%
    \put(0.4332932,0.76852216){\color[rgb]{0,0,0}\makebox(0,0)[lb]{\smash{$x$}}}%
    \put(0.34035162,0.66635049){\color[rgb]{0,0,0}\makebox(0,0)[lb]{\smash{$\vva$}}}%
    \put(0.12844491,0.84577335){\color[rgb]{0,0,0}\makebox(0,0)[lb]{\smash{$\vFLC$}}}%
    \put(0.25730112,0.81708403){\color[rgb]{0,0,0}\makebox(0,0)[lb]{\smash{$\vFDC$}}}%
    \put(0.0362889,0.67374596){\color[rgb]{0,0,0}\makebox(0,0)[lb]{\smash{$x$}}}%
    \put(0.58764427,0.80508678){\color[rgb]{0,0,0}\makebox(0,0)[lb]{\smash{$\mB$}}}%
    \put(0.19569221,0.77273815){\color[rgb]{0,0,0}\makebox(0,0)[lb]{\smash{$\mC$}}}%
    \put(0.75581025,0.90738342){\color[rgb]{0,0,0}\makebox(0,0)[lb]{\smash{$\alphaB$}}}%
    \put(0.77054398,0.80638692){\color[rgb]{0,0,0}\makebox(0,0)[lb]{\smash{$\alphazero$}}}%
    \put(0.82643768,0.84610139){\color[rgb]{0,0,0}\makebox(0,0)[lb]{\smash{$\alphad$}}}%
    \put(0.48732451,0.48956242){\color[rgb]{0,0,0}\makebox(0,0)[lb]{\smash{$\vPc$}}}%
    \put(0.39018682,0.04603483){\color[rgb]{0,0,0}\makebox(0,0)[lb]{\smash{$\mKCU$}}}%
    \put(0.77987438,0.08172308){\color[rgb]{0,0,0}\makebox(0,0)[lb]{\smash{$\vFLD$}}}%
    \put(0.94481696,0.29524161){\color[rgb]{0,0,0}\makebox(0,0)[lb]{\smash{$\alphaszero$}}}%
    \put(0.76434781,0.2878078){\color[rgb]{0,0,0}\makebox(0,0)[lb]{\smash{$\vFDD$}}}%
    \put(0.92638949,0.34916643){\color[rgb]{0,0,0}\makebox(0,0)[lb]{\smash{$\alphas$}}}%
    \put(0.71268375,0.16512753){\color[rgb]{0,0,0}\makebox(0,0)[lb]{\smash{$\mD$}}}%
    \put(0.17458739,0.38364639){\color[rgb]{0,0,0}\makebox(0,0)[lb]{\smash{$\mA$}}}%
    \put(0.62373593,0.20715764){\color[rgb]{0,0,0}\makebox(0,0)[lb]{\smash{$x$}}}%
  \end{picture}%
\endgroup%
}
\caption{Angle of attack $\alpha$, apparent air velocity $\va$, depower angle $\ad$ and $\alpha_0$ of the four point kite model. Steering is accomplished by changing $\as$. Sideslip is possible.}
\label{fig:alpha_0}
\end{figure}
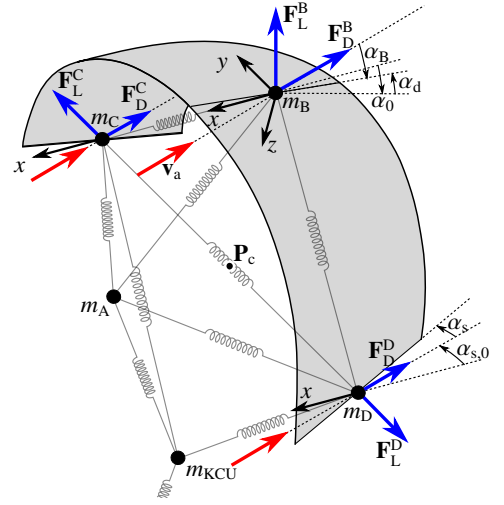

\noindent We make the following assumptions:
\begin{itemize}
\item The kite-tether angle depends linearly on the depower settings $\ud$;
\item the kite-depower angle has the value $\alpha_0$ for $\ud~=~\udd$;
\item The maximal depower value of $\ud=\udmax$ corresponds to a kite-tether angle of $\alpha_0-\admax.$
\end{itemize}
Then, the angle of attack can be calculated with the following formula
\begin{equation}
\alpha~=~\arccos \left(\frac{\vva \cdot \vex} {\va} \right) - \ad + \alpha_0 ,
\label{eq:alpha_one_point}
\end{equation}
where $\alpha_0$ is the angle between the kite and the cable when the kite is fully powered as shown in Fig. \ref{fig:alpha_0} and $\ad$ is the additional angle resulting from reeling out the depower line
\begin{equation}
\ad = \frac{\ud - \udd}{\udmax - \udd}~\admax ,
\label{eq:alpha_d}
\end{equation}
where $\udd$ is the value of the depower control input that is needed for the fully powered kite (maximal L/D) and $\udmax$ and $\admax$ the values for $\ud$ and $\ad$ respectively that are needed for the fully depowered kite.

\begin{figure}[ht]
\centering
\includegraphics[width=0.75\linewidth]{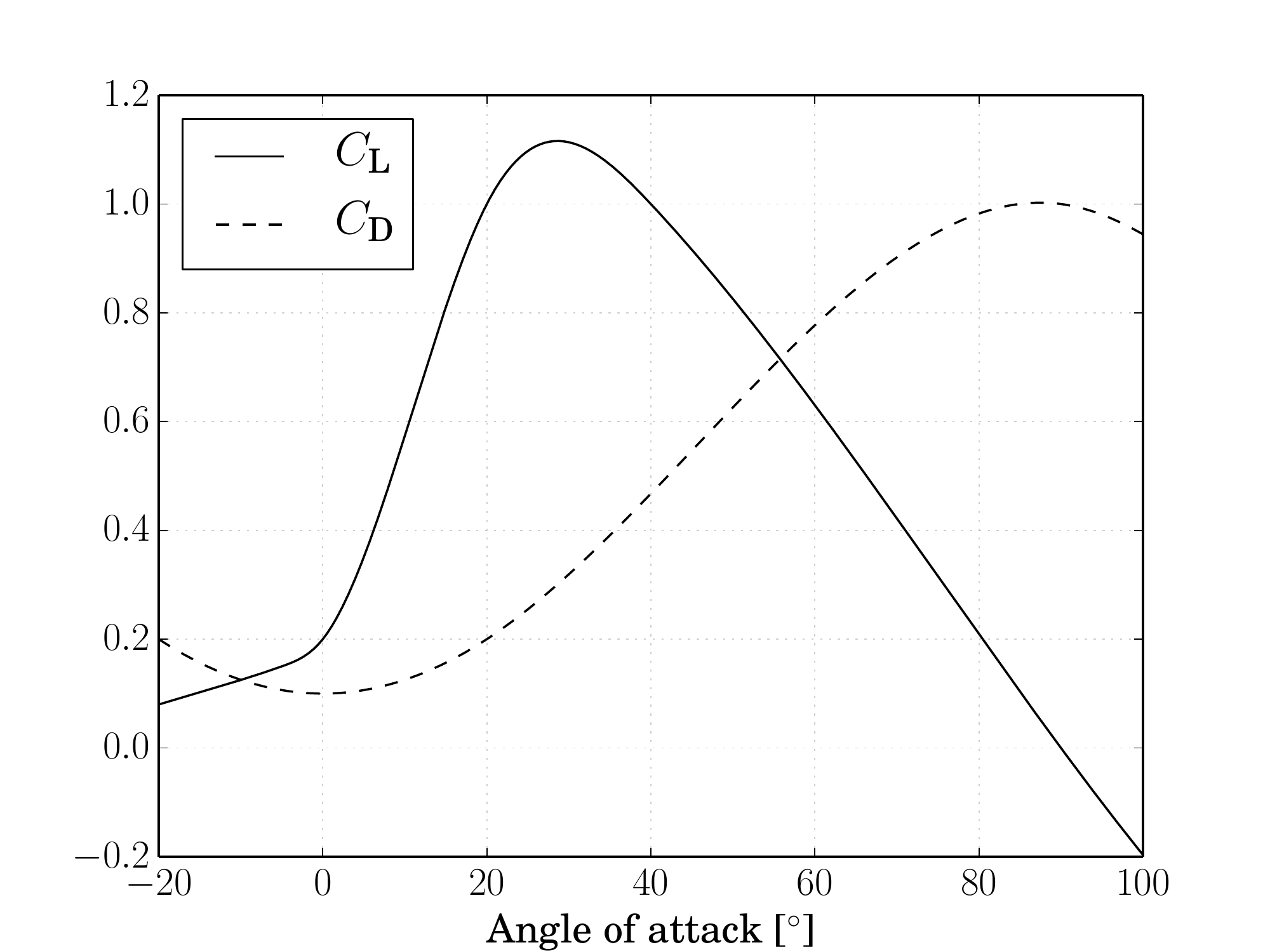}
\caption[Lift and Drag coefficients]{The lift and the drag coefficients as function of the angle of attack.}
\label{fig:lift_and_drag}
\end{figure}

Figure \ref{fig:lift_and_drag} shows the lift coefficient $\CL$ and the drag coefficient $\CD$ as functions of the angle of attack $\alpha$. The curves are established using the models of lift and drag coefficients of stalled and unstalled airfoils from \cite{Spera2008}, yet experience based modifications were made to better fit the coefficients of the non-ordinary wing section of a leading edge inflatable tube kite. 

\subsection{Four point kite model}
\label{sec:four_point_kite_model}

\noindent The point mass kite model can be sufficient to simulate and optimize the flight path of a power kite, because it is controllable during the power cycle and the simulated tether forces are close to the measured values. In addition the point mass model can be used to calculate the initial orientation of more complex models.
However, it is not a good choice for the development and optimization of flight-path control algorithms, because the reaction of the kite to steering inputs is problematic: A point mass kite has no rotational inertia, therefore its yaw angle is \emph{jumping} when the sign of $\va$ is changing. This is non-physical behaviour. In these situations controllability is lost. Therefore, we will now investigate a four-point kite model (4p model) in order to obtain a more realistic and robust model.

\subsubsection*{Geometry and mass distribution}
\begin{figure}[t]
\centering
\def\svgwidth{0.45\textwidth}
\begingroup%
  \makeatletter%
  \providecommand\color[2][]{%
    \errmessage{(Inkscape) Color is used for the text in Inkscape, but the package 'color.sty' is not loaded}%
    \renewcommand\color[2][]{}%
  }%
  \providecommand\transparent[1]{%
    \errmessage{(Inkscape) Transparency is used (non-zero) for the text in Inkscape, but the package 'transparent.sty' is not loaded}%
    \renewcommand\transparent[1]{}%
  }%
  \providecommand\rotatebox[2]{#2}%
  \ifx\svgwidth\undefined%
    \setlength{\unitlength}{1920bp}%
    \ifx\svgscale\undefined%
      \relax%
    \else%
      \setlength{\unitlength}{\unitlength * \real{\svgscale}}%
    \fi%
  \else%
    \setlength{\unitlength}{\svgwidth}%
  \fi%
  \global\let\svgwidth\undefined%
  \global\let\svgscale\undefined%
  \makeatother%
  \begin{picture}(1,0.92916667)%
    \put(0,0){\includegraphics[width=\unitlength]{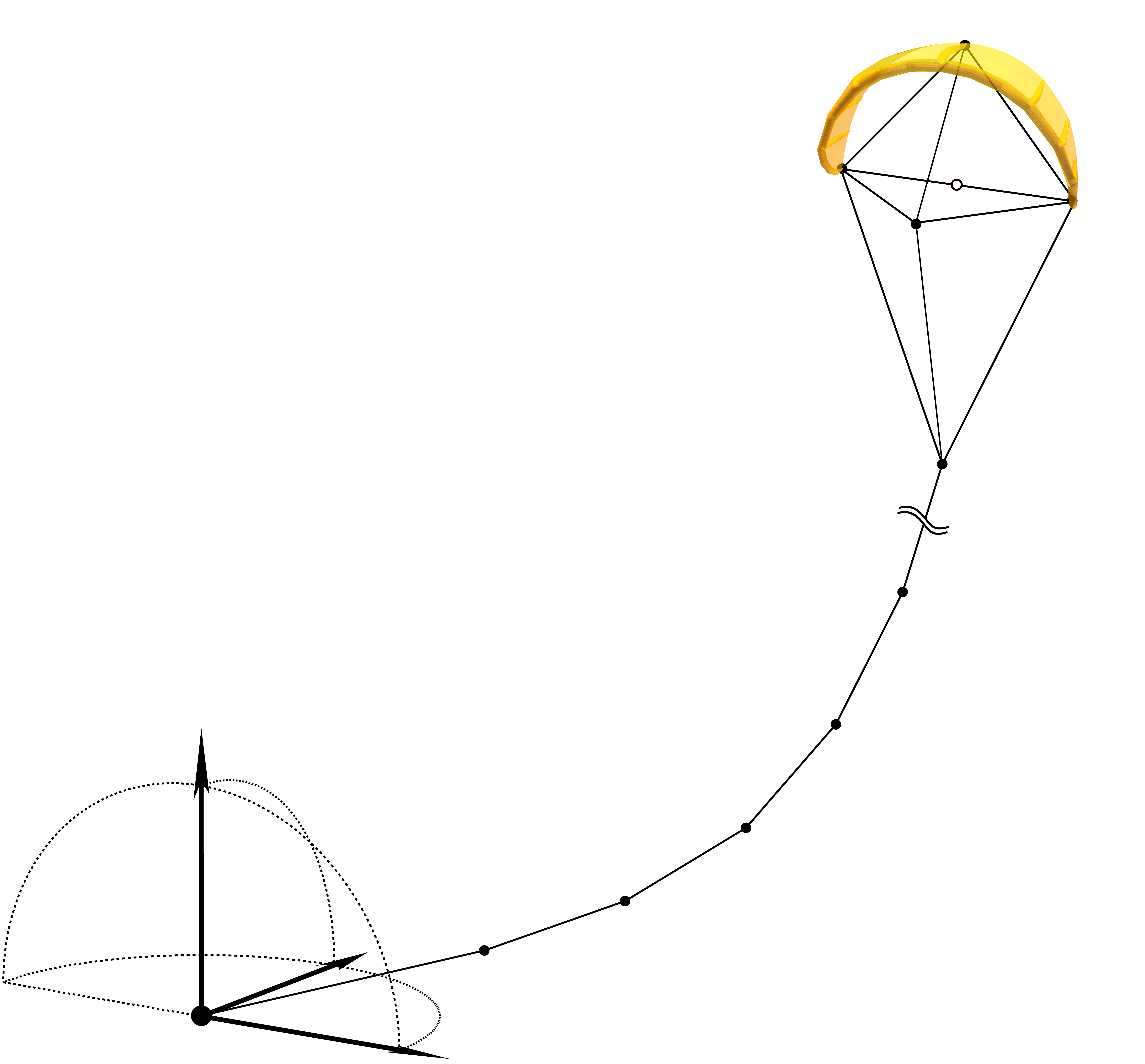}}%
    \put(0.83646118,0.90893707){\color[rgb]{0,0,0}\makebox(0,0)[lb]{\smash{$\vec{B}$}}}%
    \put(0.9484904,0.739199){\color[rgb]{0,0,0}\makebox(0,0)[lb]{\smash{$\vec{D}$}}}%
    \put(0.68812368,0.74897258){\color[rgb]{0,0,0}\makebox(0,0)[lb]{\smash{$\vec{C}$}}}%
    \put(0.81051717,0.69804108){\color[rgb]{0,0,0}\makebox(0,0)[lb]{\smash{$\vec{A}$}}}%
    \put(0.83700134,0.50594693){\color[rgb]{0,0,0}\makebox(0,0)[lb]{\smash{$\vPKCU$}}}%
    \put(0.79741557,0.3770418){\color[rgb]{0,0,0}\makebox(0,0)[lb]{\smash{$\vec{P}_5$}}}%
    \put(0.73868073,0.26231895){\color[rgb]{0,0,0}\makebox(0,0)[lb]{\smash{$\vec{P}_4$}}}%
    \put(0.65997101,0.17086036){\color[rgb]{0,0,0}\makebox(0,0)[lb]{\smash{$\vec{P}_3$}}}%
    \put(0.5535286,0.10655021){\color[rgb]{0,0,0}\makebox(0,0)[lb]{\smash{$\vec{P}_2$}}}%
    \put(0.42991496,0.06415127){\color[rgb]{0,0,0}\makebox(0,0)[lb]{\smash{$\vec{P}_1$}}}%
    \put(0.17465035,-0.00237206){\color[rgb]{0,0,0}\makebox(0,0)[lb]{\smash{$\vec{P}_0$}}}%
    \put(0.19122147,0.27742785){\color[rgb]{0,0,0}\makebox(0,0)[lb]{\smash{$\zw$}}}%
    \put(0.39659175,-0.00056181){\color[rgb]{0,0,0}\makebox(0,0)[lb]{\smash{$\xw$}}}%
    \put(0.33631983,0.11220507){\color[rgb]{0,0,0}\makebox(0,0)[lb]{\smash{$\yw$}}}%
    \put(0.84278941,0.77431226){\color[rgb]{0,0,0}\makebox(0,0)[lb]{\smash{$\vPc$}}}%
  \end{picture}%
\endgroup%
\caption{Four point model of the kite defined by points $\vec{A}$, $\vec{B}$, $\vec{C}$ and $\vec{D}$. Points $\vec{P}_0$ to $\vPKCU$ discretize the tether.} 
\label{fig:FourPointKite}
\end{figure}

\noindent The most simple particle-system based kite model that has rotational inertia in all axis is a four point kite model, which we will use from now on. The points of the this model are defined in Fig. \ref{fig:FourPointKite}.

\noindent The kite mass $\mk$ is distributed to points $\vec{A}$ to $\vec{D}$ according to Eqns. (\ref{eq:0}) to (\ref{eq:4}) while the mass of the kite control unit $\mKCU$ plus half of the mass of the last tether segment are used as the mass~of~$\vPKCU$
\begin{align}
m_{\mr{PKCU}} &= \mKCU + \frac{\lt~\sigma}{2~n} , \label{eq:0}\\
\mA &= \gamma ~ \mk , \label{eq:1}\\
\mB &= 0.4 ~ (1 - \gamma) ~ \mk , \label{eq:2}\\
\mC &= 0.3 ~ (1 - \gamma) ~ \mk , \label{eq:3}\\
\mD &= 0.3 ~ (1 - \gamma) ~ \mk , \label{eq:4}
\end{align}
where $\gamma$ is the nose mass fraction of the wing, $n$ the number of tether segments, $\lt$ the current tether length and $\sigma$ the linear mass density of the tether. The simulation of typical flight manoeuvres at low apparent air velocities has shown that a value of $\gamma = 0.47$ reproduces well the dive-down behaviour of the Leading Edge Inflatable (LEI) tube kites employed in the current study.

The virtual centre position of the kite, $\vec{\vPc}$ is defined as 
\begin{equation}
\vec{\vPc}~=~\frac{1}{2}~(\vec{C} + \vec{D}) .
\end{equation}
The origin of the kite reference frame is at $\vec{B}$. The unit vectors $\vex, \vey$ and $\vez$ are defined as
\begin{align}
\vez &=~\frac{\vec{\vPc} - \vec{B}}{\parallel \vec{\vPc} - \vec{B} \parallel} , \label{eq:x_4point} \\ 
\vey &=~\frac{\vec{C} - \vec{D}}{\parallel \vec{C} - \vec{D} \parallel} , \label{eq:y_4point} \\
\vex     &=~\vey \times \vez . \label{eq:z_4point}
\end{align}
To parametrize the shape of the kite only three values need to be defined: The height of the kite $\hk$ (distance between $\vPc$ and $\vec{B}$), the height of the bridle $\hb$ (distance between $\vPc$ and $\vPKCU$) and the width of the kite $\wk$ (the distance between $\vec{C}$ and $\vec{D}$).

\subsubsection*{Initial conditions}

To calculate the initial positions of the kite particles, the point mass kite model from Sect. \ref{subsec:point-mass_model} is used. The initial unit vectors of the kite reference frame ($\vexx, \veyy$ and $\vezz$) are calculated using the kite position, the orientation of the last tether segment and the apparent air velocity (Eqns. (\ref{eq:z_vector}), (\ref{eq:y_vector}) and (\ref{eq:x_vector})).

When these vectors are known, the positions of the kite particles at zero force can be defined by the following equations
\begin{align}
\vec{\vPc} &= \vPKCU - \hb ~ \vezz , \\
\vec{A} ~&= \vec{\vPc} + \dnr ~ \wk ~\wrel~ \vexx , \\
\vec{B} ~&= \vec{\vPc} - h_\mr{k} ~ \vezz , \\
\vec{C} ~&= \vec{\vPc} + 0.5  ~ \wk ~\wrel~ \veyy , \\
\vec{D} ~&= \vec{\vPc} - 0.5  ~ \wk ~\wrel~ \veyy ,
\end{align}
where $\dnr$ is the relative nose distance, a kite dependant factor in the order of $0.2$. In combination with the nose mass fraction $\gamma$ the factor $\dnr$ can be used to tune the rotational inertia and the centre of gravity. The distance from $\vec{C}$ to $\vec{D}$ is calculated using the tip-to-tip distance of the kite $\wk$ multiplied with the relative kite width $\wrel$ which is a factor in the order of $0.9$ and reflects the fact that the aerodynamic steering forces do not act on the tips of the kite, but a little bit further inwards.

During the simulation, the aerodynamic forces are applied to surfaces that are attached to the kite particles. This causes them to change their positions, and from the current positions the unit vectors of the kite reference frame can then be calculated using Eqns. (\ref{eq:x_4point}), (\ref{eq:y_4point}) and (\ref{eq:z_4point}). 

\subsubsection*{Projected air velocities and angles of attack}
\noindent The aerodynamic model assumes surfaces attached to the top particle $\vec{B}$ and to the side particles $\vec{C}$ and $\vec{D}$. The sole purpose of particle $\vec{A}$ is to achieve rotational inertia and to realistically place the centre of gravity, therefore no aerodynamic force is attached to this particle.

The lift forces are determined based on the part of the apparent velocity that is perpendicular to the leading edge as suggested in \cite{Obert2009}.
For the surface attached to the top particle, this is the apparent velocity in the $xz$-plane $\vvaxz$. For the surfaces attached to the side particles, the apparent velocity in the $xy$-plane $\vvaxy$ is needed. These can be calculated as follows
\begin{align}
\vvaxz &= \vva - (\vva \cdot \hspace{2pt}  \vey)~\vey , \\
\vvaxy &= \vva - (\vva \cdot \hspace{2pt}  \vez)~\vez .
\end{align}
For the top surface of the kite the angle of attack can be calculated as follows
\begin{equation}
    \alphaB = \pi -\arccos ~ \left( \frac{\vec{v}_{\mr{a,B},xz} \cdot \vex}{\parallel \vec{v}_{\mr{a,B},xz} \parallel} \right)- \ad + \alpha_0 .
\end{equation}
The angle $\ad$ is the change of the angle between the kite and the last tether segment due to the change of the depower settings. The value of $\ad$ is between zero when fully powered and - for the leading edge inflatable tube kites used at Delft University of Technology - about \ang{30} when fully depowered. If the reel-out length of the depower tape, the height of the bridle, the height of the kite and the power-to-steering-line distance are known, $\ad$ can be calculated geometrically; In many cases the  linear approximation given by Eq.~(\ref{eq:alpha_d}) is sufficient.

For the side surfaces of the kite the angles of attack can be calculated as follows
\begin{align}
    \alphaC = \pi - \arccos  \left(\frac{\vec{v}_{\mr{a,C},xy} \cdot \vex}{\parallel \vec{v}_{\mr{a,C},xy} \parallel} \right)- \as + \alphaszero , \\
    \alphaD = \pi - \arccos  \left(\frac{\vec{v}_{\mr{a,D},xy} \cdot \vex}{\parallel \vec{v}_{\mr{a,D},xy} \parallel} \right)+ \as + \alphaszero ,
\end{align}
where $\as$ is the change of the angle of attack caused by the steering line difference. For $\alphaszero$ a value of $10\,^{\circ}$ is assumed. With $\uss$ a steering offset - which is in practice unavoidable and caused by asymmetries in the steering system - it can be calculated as follows
\begin{equation}
\as = \frac{\us - \uss}{1+\kds(\ad / \alphadmax)}  ~ \alphasmax .
\end{equation}
The value of $\alphasmax$ (in the order of \ang{20} must be chosen such that the steering sensitivity of the kite model matches the steering sensitivity of the kite to be simulated.
The factor $\kds$ describes the influence of the depower angle $\ad$ on the steering sensitivity: depending on the geometry of the bridle it has a value in the range of $1<\kds<2$. A value of $1.5$ means that the fully depowered kite needs $2.5$ times the steering input as a fully powered kite to achieve the same turn rate (under the condition that the apparent wind speed is the same). 

\subsubsection*{Aerodynamic forces}
\label{subsec:aero_forces}
\noindent Steering is accomplished by changing the angle of attack for the side surfaces differentially.
The aerodynamic forces that act on $\vec{B}$, $\vec{C}$ and $\vec{D}$ can be calculated according to Eqns. (\ref{eq:lp2}) to (\ref{eq:dp4}), where $\Aside/A$ is the relative side area of the kite and $\rho$ the air density.

\begin{align}
\vFLB &= \frac{1}{2}~\rho~\vaBxz^2~ A~\CL(\alphaB)~ \frac{\vec{v}_\mr{a,B} \times \vey}{\parallel \vec{v}_\mr{a,B} \times \vey \parallel} , \label{eq:lp2}\\
\vFLC &= \frac{1}{2}~\rho~ \vaCxy^2~ A\frac{\Aside}{A}~\CL(\alphaC)~ \frac{\vec{v}_\mr{a,C} \times \vez}{\parallel \vec{v}_\mr{a,C} \times \vez \parallel} , \\
\vFLD &= \frac{1}{2}~\rho~ \vaDxy^2~ A\frac{\Aside}{A}~\CL(\alphaD)~ \frac{\vez \times \vec{v}_\mr{a,D} }{\parallel \vez \times \vec{v}_\mr{a,D} \parallel} , \\
\vFDB &= \frac{1}{2}~\rho~K_\mr{D}~ \vaB^2~ A~\CD(\alphaB)~\frac{\vec{v}_\mr{a,B}}{\parallel \vec{v}_\mr{a,B} \parallel} , \\
\vFDC &= \frac{1}{2}~\rho~K_\mr{D}~ \vaC^2~ A\frac{\Aside}{A}~\CD(\alphaC)~\frac{\vec{v}_\mr{a,C}}{\parallel \vec{v}_\mr{a,C} \parallel} , \\
\vFDD &= \frac{1}{2}~\rho~K_\mr{D}~ \vaD^2~ A\frac{\Aside}{A}~\CD(\alphaD)~\frac{\vec{v}_\mr{a,D}}{\parallel \vec{v}_\mr{a,D} \parallel} .
\label{eq:dp4}
\end{align}
The coefficient $K_\mr{D}$ is required to achieve the same lift-to-drag ratio for the straight flying four point kite as for the one point kite. It can be calculated from
\begin{equation}
K_\mr{D} = \left(1 -\frac{\Aside}{A}\right)~\kappa 
\label{eq:k}
\end{equation}
where $\kappa=0.93$ was needed to compensate the higher drag coefficients of the side areas, compared to the top area, caused by~$\alphaszero$.

\subsection{Winch model}
\label{subsec:winch}
\noindent We view the winch as the assembly of an asynchronous generator, a gearbox and a drum around which the tether is wound. The generator is used as motor during the reel-in phase and the sign of the generator's torque determines the direction of the energy flow. We modelled the winch by combining the differential equations for the inertial system and an expression for the torque-speed characteristics of the generator.

\subsubsection*{Inertial dynamics of the winch}

\noindent The differential equations for the winch are again defined as an implicit problem
\begin{align}
    F(t,\ye,\dotye)~&=~0 , \label{eq:main_equation_winch}\\
\ye(t_0)~&=~\yezero , \\
\dotye(t_0)~&=~\dotyezero .
\end{align}
The vector $\ye$ is the extended state vector of the implicit problem and consists of tether length $\lti$ and the tether velocity $\vto$
\begin{equation}
\ye ~=~ (\lti, \vto) .
\end{equation}

\begin{sloppypar}
    \noindent In order to solve this problem the residual $\rre~=~F(t,\ye, \dotye)$ is to be calculated, with $\re$ defined as
\end{sloppypar}
\begin{equation}
	\rre =
	\begin{bmatrix}
		\vto - \dotlti \\
		a_{t,o} - \dotvto
	\end{bmatrix} .
\end{equation}

Here, $a_{t,o}$ is the acceleration of the tether at the ground station. Under the assumption of an inelastic interconnection of the generator and drum through the gearbox, the acceleration can be calculated as
\begin{equation}
\ato = \frac{1}{I}~\frac{r}{n}~\left( \taug + \taud - \tauf \right) ,
\end{equation}
where $I$ is winch inertia as seen from the generator, $r$ the drum radius, $n$ the gearbox ratio, $\taug$ the generator torque, $\taud$ torque exerted by the drum on the generator and $\tauf$ the friction torque.

The torque exerted by the drum depends on the tether force that is exerted on the drum, which equals the norm of the force on the first tether particle
\begin{equation}
\taud = \frac{r}{n}~\|~\boldsymbol{f_s}_{,0}~\| .
\end{equation}
We modelled the friction as the combination of a viscous friction component with friction coefficient $\cf$ and static friction $\taus$
\begin{equation}
\tauf = \cf~\vto + \taus~\mr{sign}(\vto) .
\end{equation}
%
%
%

\subsubsection{Torque profile of the asynchronous generator}
\noindent To determine the torque-speed profile of the asynchronous generator, we used the equivalent circuit representation as in~\cite[p. 326]{Wildi2002}. Under the assumption of negligible stator resistance, $\taum$ can be expressed as a function of $\vto$ and the synchronous generator speed $\vs$ as
\begin{equation}
\taug = \alpha~\frac{\vs - \vto}{1 + \beta~(\vs - \vto)^2} .
\end{equation}
We assumed that the generator voltage $E$ is increasing linearly with the set speed, up to the nominal voltage $\en$ at the nominal synchronous speed $\vsn$ of the generator
\begin{equation}
	E = 
	\begin{dcases}
		\en \frac{\vs	}{\vsn}	& \text{if } |~ \vs ~| \leq \vsn \\
		\en						& \text{if } |~ \vs ~| > \vsn
	\end{dcases} .
\end{equation}
As derived in \cite{Schreuders2013}, the parameters $\alpha$ and $\beta$ can be expressed as
\begin{align}
	\alpha &=
	\begin{dcases}
		\frac{\en^2~r}{\vsn^2~\rrr~n}		& \text{if } |~ \vs ~| \leq \vsn \\
		\frac{\en^2~r}{\vs^2~\rrr~n}	& \text{if } |~ \vs ~| > \vsn
	\end{dcases} \\
	\beta &= \frac{L^2}{\rrr^2}~\frac{n^2}{r^2} ,
\end{align}
where $\rrr$ is the rotor resistance and $L$ is the generator's self inductance. These generator parameters could either be measured or estimated based on known torque data.

\begin{table}[t]
\centering
\caption{Properties of the ground station of Delft University of Technology}
\label{tab:winc_properties}
\begin{tabular}{ l r l }
\textbf{Ground station} & ~ & ~ \\ 
Gearbox ratio $n$ [-] & 6.2 & ~ \\ 
Drum radius $r$ [m] & 0.1615& \\ 
Inertia $I$ [kg m$^2$]& 0.328 & ~ \\ 
Viscous friction coeff. $\cf$ [Ns]& 0.799 & ~ \\ 
Static friction $\taus$ [Nm] & 3.18 & ~ \\  
Rotor resistance $\rrr$ [m$\Omega$] & 72.7 & ~ \\ 
Self inductance $L$ [mH] & 2.97 & ~ \\ 
Nominal synchronous speed $v_{s,n}$ [m/s]& 4.09& ~ \\
Nominal voltage $\en$ [V] & 231 & ~
\end{tabular} 
\end{table}

\subsection{Control system}
In this section a brief description of the control system is given. Further details can be found in \cite{Vlugt2013} and \cite{Fechner2012}.
\subsubsection*{Flight path planning and control}
For the automated power production a simple flight path planner is used: The kite is always steered towards one of three points: During reel-in and parking it is steered towards zenith (directly above the ground station). During reel-out it is steered to a point on either the right or left side of the wind window~\cite{Vlugt2013}.

The orientation of the kite (the heading angle) is controlled. Great circle navigation is used to determine the heading needed to steer towards the target point. The difference between the required heading and the actual heading is the error signal that is going into a PI controller that is controlling the steering signal $\is$ of the kite control unit. In addition the KCU has an input $\id$ for the depower signal. The set value $\id$ is low during reel-out and high during reel-in (predefined, fixed values).

The steering signal differentially changes the length of the left and right steering lines, the depower signal changes the length of the steering lines relative to the length of the depower lines. The actuators are modelled such that they have a maximum speed (derivative of the output control signals $\us$ and $\ud$). They use a P-controller to control the output signal. In addition a delay of 150~ms was implemented in the model. The delay is mainly caused by the motor controllers.
\subsubsection*{Winch control}
During reel-out the winch is using a set value for the reel-out speed in addition to a maximal value of the tether force. The speed is used as long as the maximum tether force is not exceeding the set value, otherwise the synchronous speed is increased to limit the force. A parameter varying PID controller is used to track the set values.

During reel-in, different values for the set force and set speed are used. Soft transitions are implemented for the set values when switching between reel-in and reel-out.

\subsection{Implementation and accuracy}
The Radau5DAE solver \cite{Hairer1996} from version 2.4 of the Assimulo suite \cite{Anderson2013} is used for solving the differential algebraic system, as it offered the best performance.
\subsubsection*{Real-time simulation based on the numerical model}
\label{sec:real_time_simulation}
Because for software-in-the-loop testing of kite control components a batch simulation is not sufficient, a soft real time simulator was implemented. The real-time simulation is executed in the following way: After the start of the simulation a new system state is calculated in fixed time intervals of currently 50~ms. The new state is then published and used by the KPS controller to calculate new values for steering and depower settings of the kite and for the set-value of the reel-out speed of the winch. These values are assumed to be constant during the next time interval. Within one simulation time interval, the implicit equation system solver uses as many time-steps as necessary to calculate a solution with the specified precision.

\subsubsection*{Model and measurement accuracy}
The solver that was used allows it to specify a maximum error. This error was set to 1.8~cm for the position states and to 0.03 cm/s for the velocity states. The tether was discretised with seven particles. 

\noindent The wind sensor at the ground has an accuracy of 5\% plus 0.5 knots. The tether force was measured with an accuracy of 1\% $\pm$ 10~N, the reel-out speed with 2\% $\pm$ 0.05 m/s.

\newcommand{\specialcell}[2][r]{%
  \begin{tabular}[#1]{@{}l@{}}#2\end{tabular}}
  
\renewcommand{\arraystretch}{1.2}


\section{Model calibration and results}
\label{sec:model_calibration}
\noindent For the calibration of the model the following steps are needed:
\begin{enumerate}
\item determine the physical system properties (Table \ref{tab:kite_properties}) and enter them as parameters into the model
\item determine the wind profile;
\item determine the lift-over-drag ratio of the kite as function of the depower settings;
\item determine the steering coefficients of the kite;
\item validate the average and maximum force during reel-out;
\item validate power output over the full cycle.
\end{enumerate}
The one-point model, the four-point model and the HYDRA kite of Delft University of Technology are compared. The models were tuned to match the kite properties as much as possible.

\subsection{Test flight}
\noindent For parameter fitting and validation the measurements of a test flight were chosen, that took place at the Maasvlakte II, The Netherlands on 23 June 2012. The wind was very strong and the wind profile was expected to be similar to offshore conditions. This flight was chosen because it contains different flight manoeuvres, e.g. parking the kite at zenith at different heights and with different depower settings. This allows for a partial validation of the lift-over-drag properties of the kite as function of the depower settings.
\begin{figure}[ht]
\centering
\includegraphics[width=0.9\linewidth]{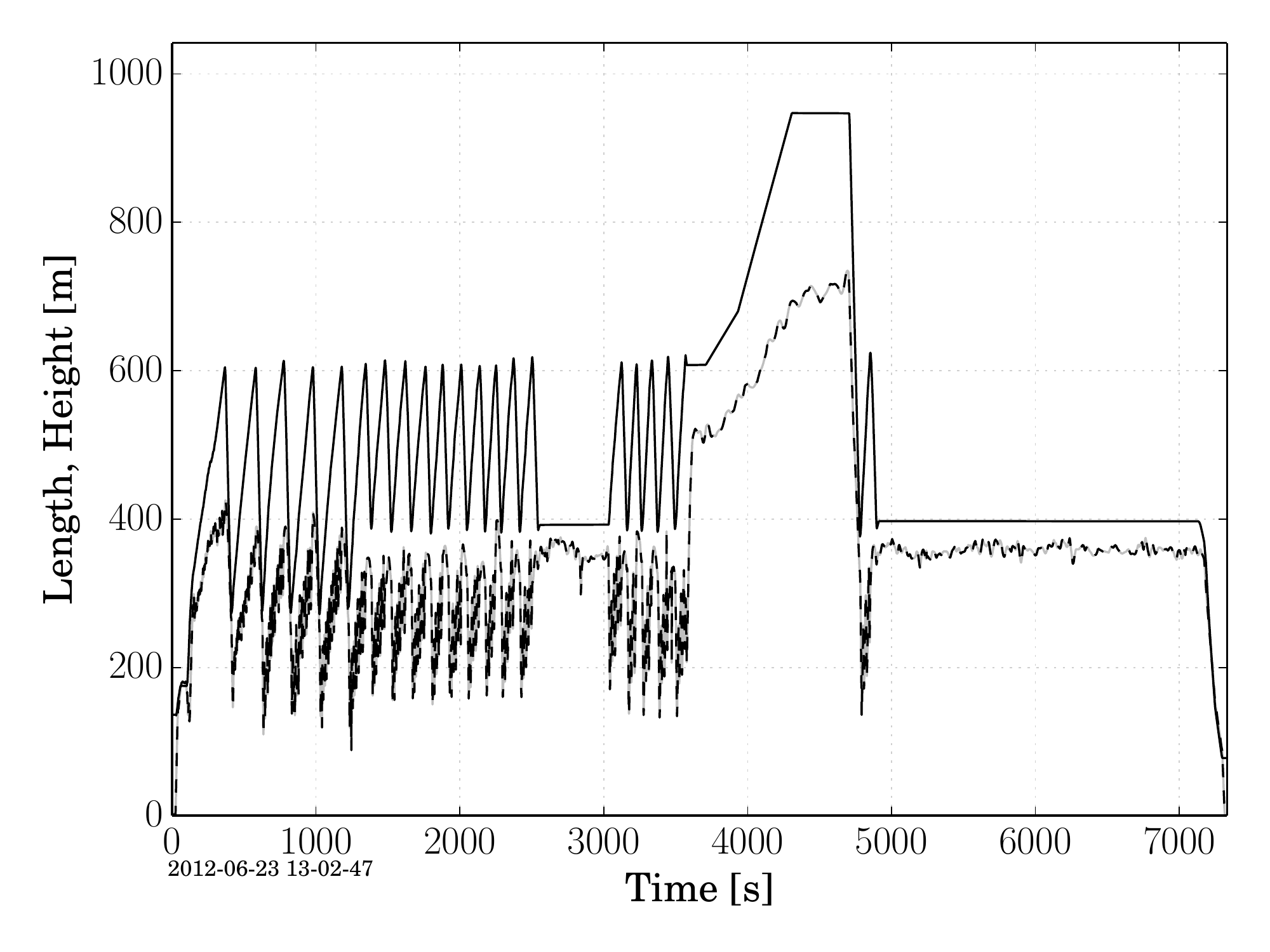}
\caption{Measured reel-out tether length $\lt$ (solid) and height $\zk$ (dashed) of the kite during a test flight on 23 June 2012 at the Maasvlakte II, The Netherlands.}
\label{fig:test_flight}
\end{figure} 
\begin{table}[t]
\centering
\caption{Properties of the HYDRA kite, bridle, KCU and tether of Delft University of Technology}
\label{tab:kite_properties}
\begin{tabular}{ l r l }
\textbf{Kite} & ~ & ~ \\ 
Projected wing surface area $A$ [m$^2$] & 10.18 & ~ \\ 
Mass including sensors $\mk$ [kg] & ~~6.21& \\ 
Width $\wk$ [m]& 5.77 & ~ \\ 
Height $\hk$ [m]& 2.23 & ~ \\ 
Relative side area $\Aside/A$ [\%] & 30.6 & ~ \\ 
\hline \textbf{Bridle} & ~ & ~ \\ 
Height $\hb$ [m]& 4.9 & ~ \\ 
Bridle line diameter [mm]& 2.5 & ~ \\ 
\hline \textbf{Kite Control Unit} & ~ & ~ \\ 
Mass $\mKCU$ [kg]& 8.4 & ~ \\ 
\hline \textbf{Main Tether} & ~ & ~ \\ 
Diameter $\dt$ [mm]& 4.0 & ~ \\
Mass per m [kg/m]& 0.013 & ~ \\ 
Unit damping coefficient $c_0$ [Ns]& 473 & ~ \\ 
Unit spring constant $k_0$ [N]& 614600 & ~ \\ 
\end{tabular} 
\end{table}
\subsection{Parking manoeuvres for aerodynamic measurements}
\label{sub:paramter_identification}
\noindent The lift-over-drag ratio and the wind profile were determined by keeping the kite pointing towards the small-earth zenith without reeling in or out. Subsequently, we waited until a force equilibrium was reached. In this situation the elevation angle of the tether is depending mainly on the lift-over-drag value, and the tether force is mainly depending on the wind speed at the height of the kite.

The measurements of Table \ref{tab:ForcesAndElevations} were used to calibrate the L/D of the kite and the sensitivity to changes of $\ud$ by changing  $\udd$ and $\alphadmax$ (see Eq.~(\ref{eq:alpha_d})).

In addition, this data was used to tune the wind profile coefficients $u_{z,0}$ and $K$ according to 
Eq.~(\ref{eq:windProfile}). The parameters $u_{z,0}, \udd, \alphadmax, K$ and $c_{d,t}$ were fitted until the force and the elevation angle for all three measurements matched with an error of less than one $\pm \sigma$. The results are shown in Table \ref{tab:simulation_parameters}
\begin{table}[]
\centering
\caption{Identified system parameters}
\label{tab:simulation_parameters}
\begin{tabular}{ l r l }
\multicolumn{3}{l}{\textbf{Fitted parameters}}\\ 
\hline $\udd$ [\%] & 21.3 & depower offset \\ 
$z_0$ [m] & 2.0e-4 & surface roughness \\ 
$K$ [-] & 1.0& wind profile correction \\ 
$\alphadmax$ [$^o$]& 31.00 & max. depower angle \\ 
$\cdt$ & 0.96 & tether drag coefficient \\ 
\multicolumn{3}{l}{\textbf{Measured parameters}}\\ 
\hline
$\umax$ [\%]& 42.47 & max. depower setting \\ 
\end{tabular} 
\end{table}
and the resulting wind profile in Fig. \ref{fig:wind_profile}. The value of $\alphadmax$ is very close to the geometrically derived value of about 30$^o$. The tether drag coefficient is very close to the value of about 1.0, that was suggested in \cite[p. 253]{Fechner2013}.

\begin{table*}[t]
\centering
\caption{Forces and elevations $\beta$ while parking}
\label{tab:ForcesAndElevations}
\begin{tabular}{ c c c c c c c c }
    Test case & \textbf{$\vwg~[ms^{-1}]$} & $\lt~[m]$ &\textbf{$\ud$}& Force [N] & $\sigmaf$ & $\beta$ [$ ^{\circ}$] & $\sigma_{\beta}$  \\ 
\hline Parking 392a & 10.35 & 392.0 & 25.1\% & 850.5 & 309.8 & 65.9 & 2.0  \\ 
Parking 392b & ~~9.59& 392.0 & 27.9\% & 551.3 & 125.1 & 60.6 & 0.9 \\ 
Parking 947~~~& 10.02 & 947.2 & 28.0\% & 552.8 &~~57.2 & 49.3 & 0.9 \\ 
\end{tabular} 
\end{table*} 

\subsection{Identifying the steering sensitivity parameters}
\noindent According to \cite[p. 149]{Erhard2013} the turn rate of the kite around the straight line between the kite and the tether should depend on the steering input $\as$, the apparent air velocity $\va$, the elevation angle $\beta$ and the orientation of the kite $\psi$ in following way
\begin{equation}
\dot{\psi}~=~c_1~\va~(\us - c_0)~+~\frac{c_2}{\va}~\sin\psi~\cos\beta .
\label{eq:psi}
\end{equation}
We added the steering offset $c_0$, because it had a relevant effect in our flight tests.

To fit the parameter $c_2$ the relative kite width $\wrel$ was varied and to fit $c_1$ the maximal steering angle $ \alphasmax$ until the measured values $c_1$ and $c_2$ matched the simulated values within 1\%.

\begin{figure*}[htbp]
\centering
\begin{subfigure}{.49\textwidth}
    \centering
    \includegraphics[height = 6.51cm, width=\linewidth]{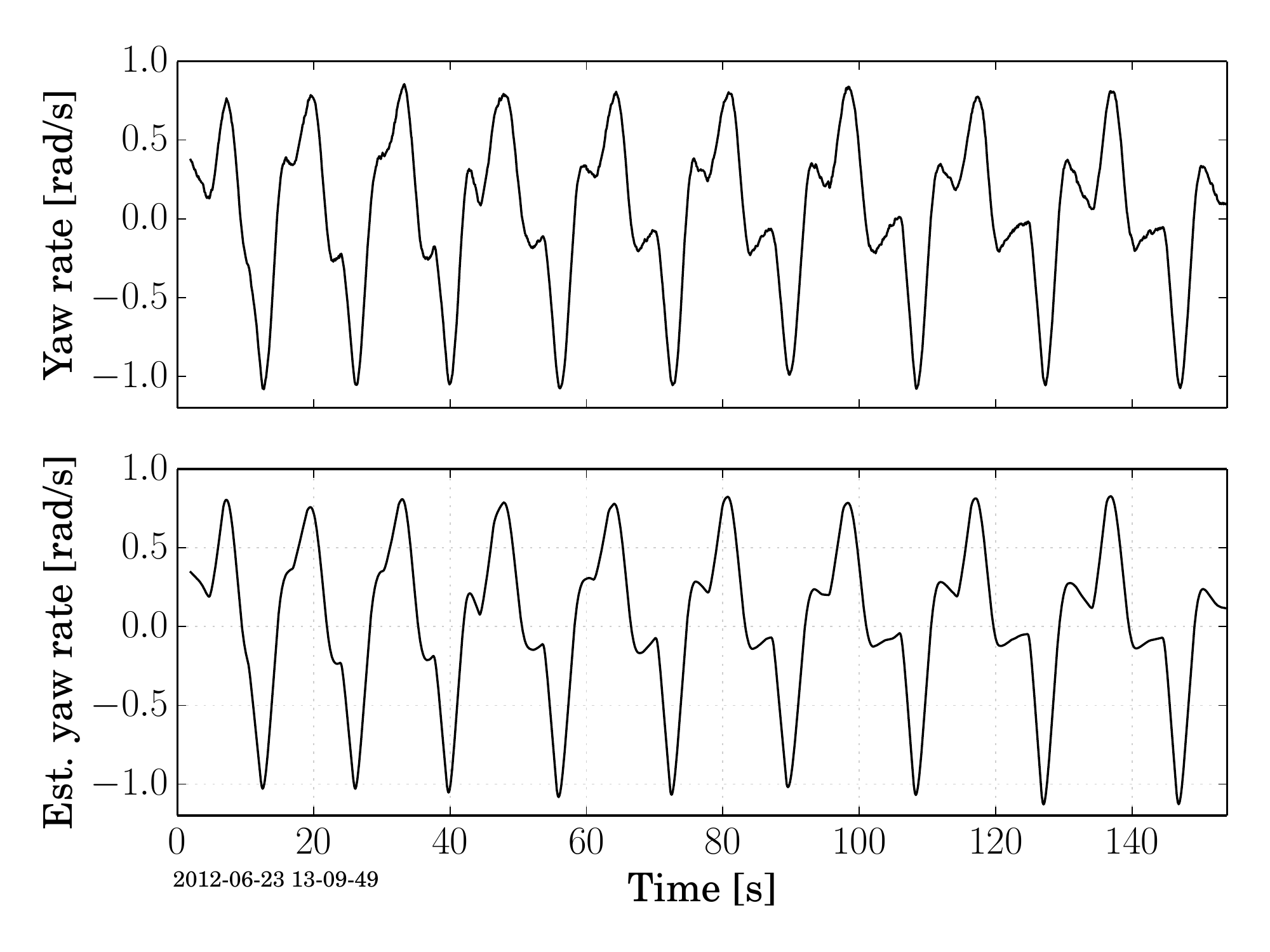}
    \caption{Measured and estimated yaw rate. The reason for the negative peaks, compared to the simulation are probably measurement errors.} 
\end{subfigure}
\hfill
\begin{subfigure}{.49\textwidth}
    \centering
    \includegraphics[height = 6.51cm, width=\linewidth]{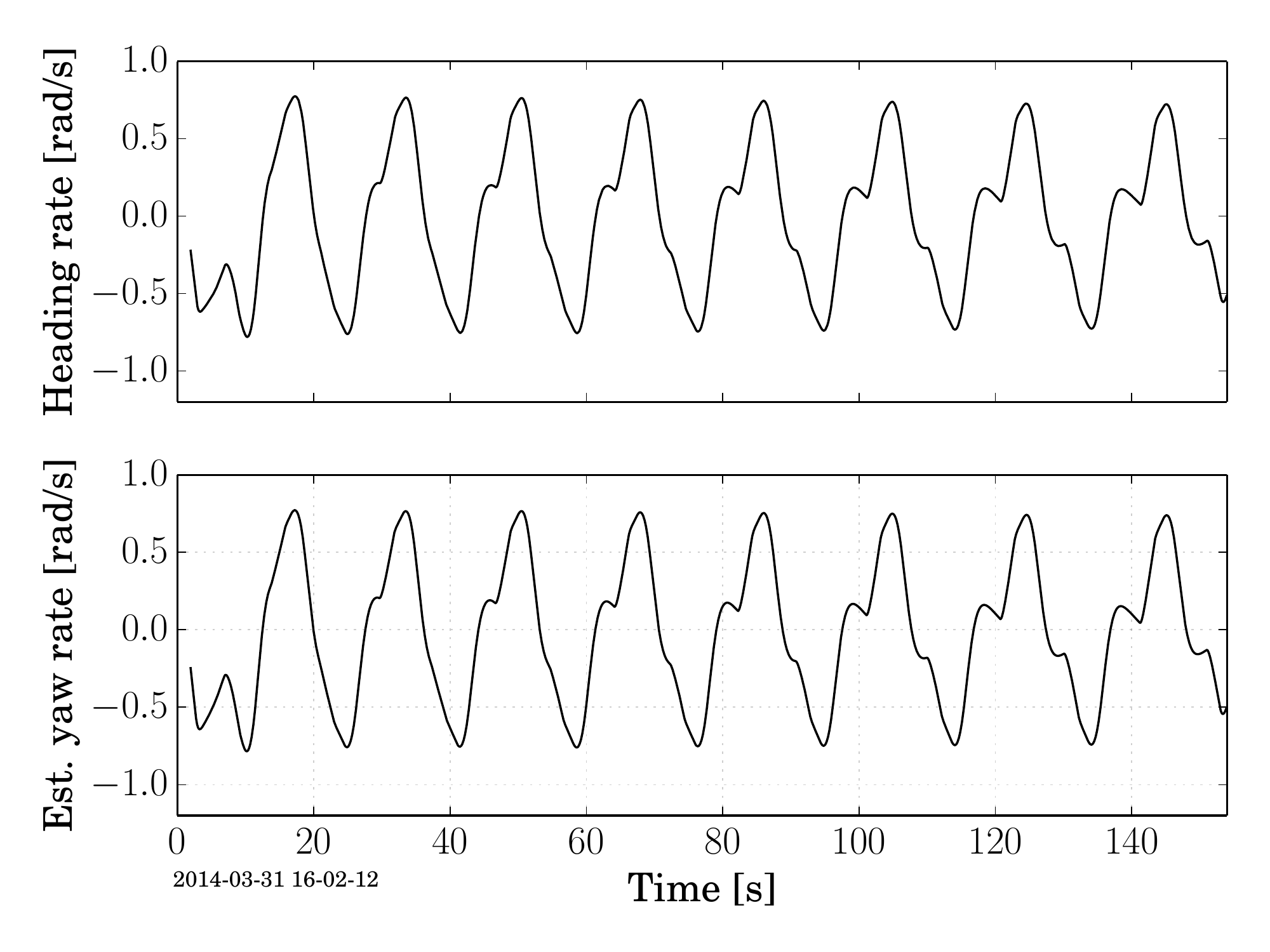}
    \caption{Simulated and estimated heading rate. The estimation is based on the turn rate law (Eq. \ref{eq:psi}), using the fitted parameters $c_0, c_1$ and $c_2$.}
\end{subfigure}
\\
\begin{subfigure}{.49\textwidth}
  \centering
  \includegraphics[height = 6.51cm, width=1.08\linewidth]{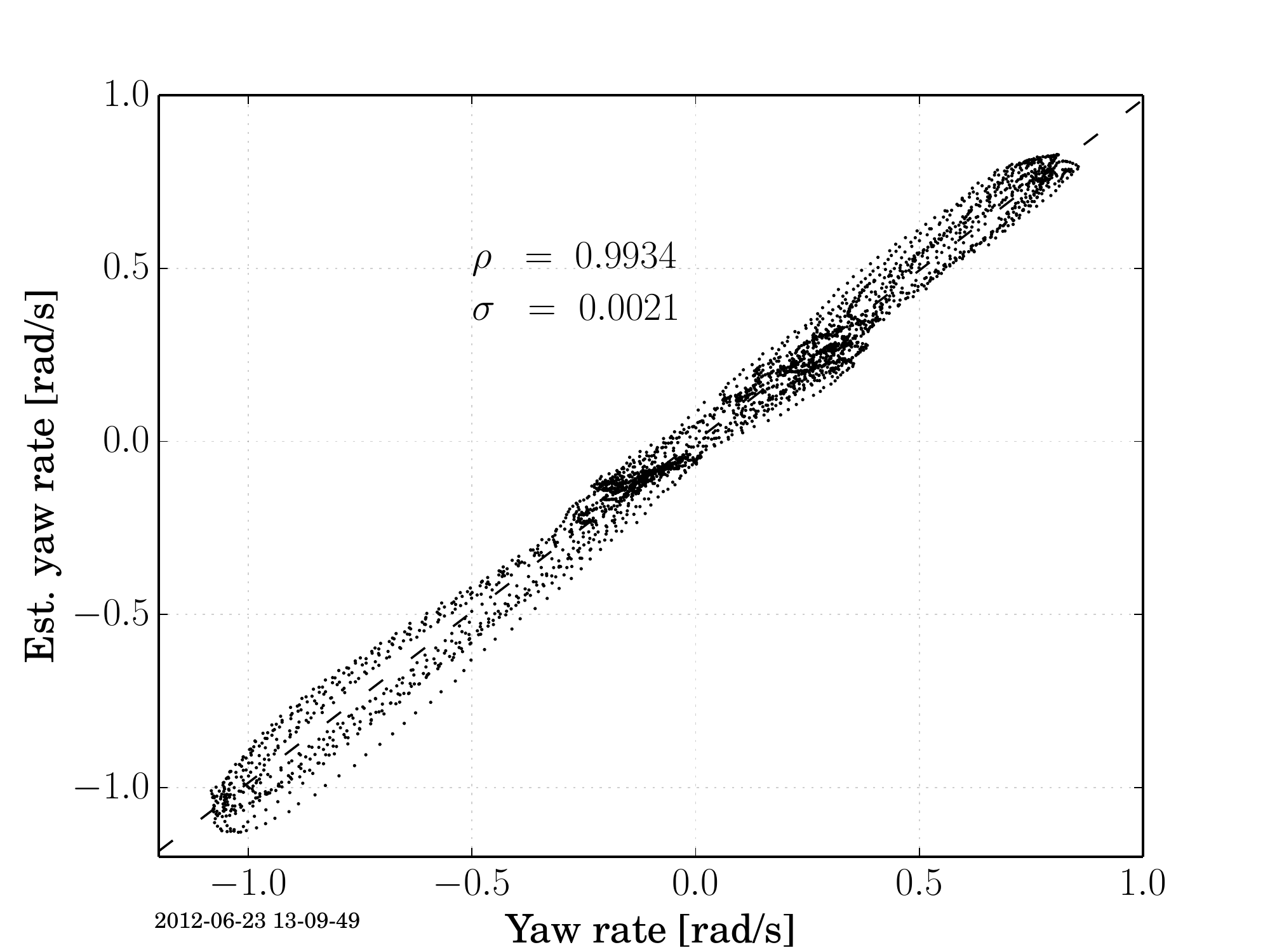}
  \caption{Estimated vs. measured yaw rate. The diagram shows a good match of the estimation with the measurement.}
\end{subfigure}
\hfill
\begin{subfigure}{.49\textwidth}
  \centering
  \includegraphics[height = 6.51cm, width=1.08\linewidth]{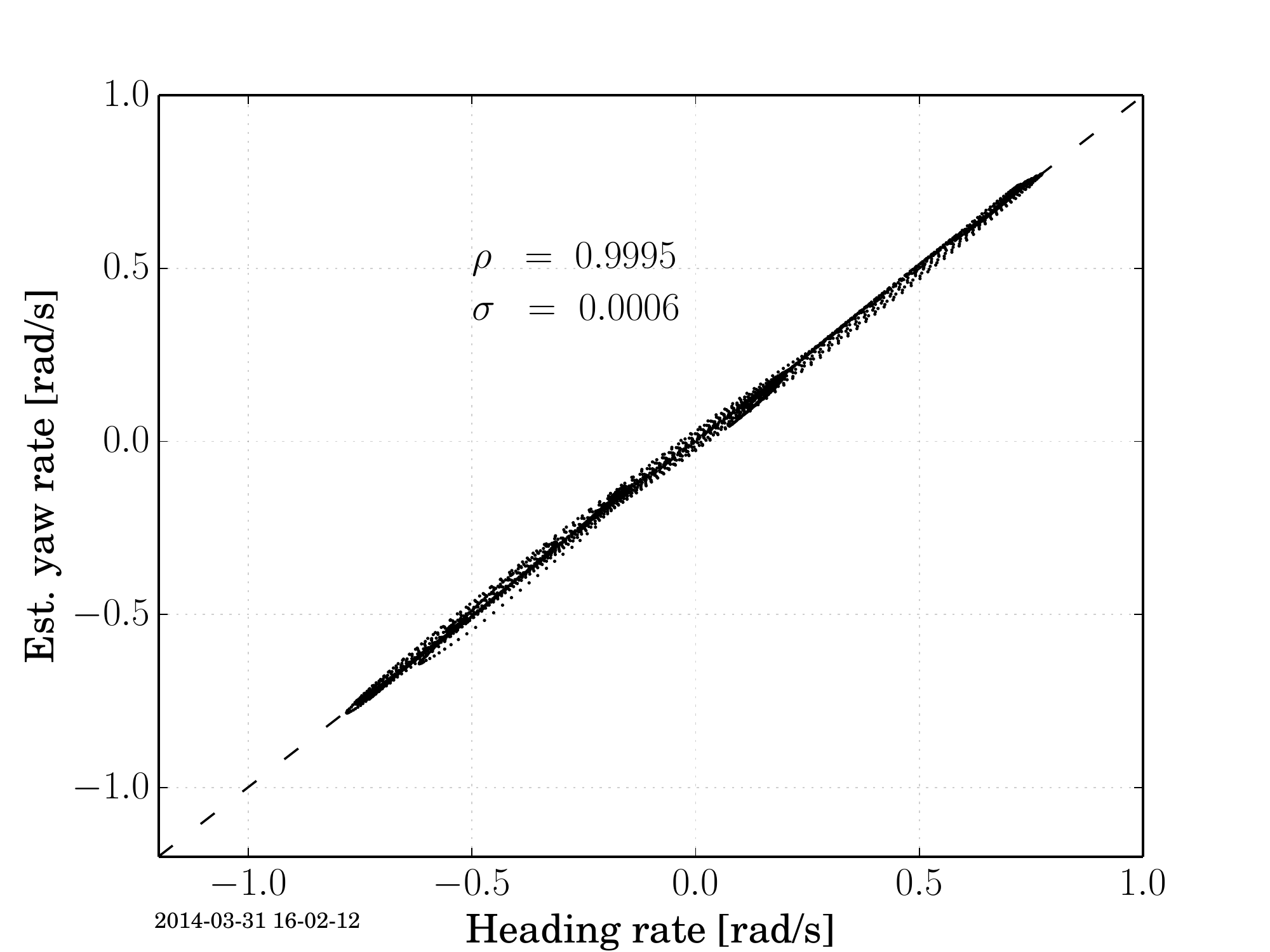}
  \caption{Estimated vs. simulated heading rate. The diagram shows a very good match of the turn rate law and the dynamic simulation.}
\end{subfigure}
\caption{Accuracy of the turn rate law applied to the measured and simulated reel-out phase of the kite. Only the simulation results of the four point model are shown, because the results for the point mass model look very much the same. $\rho$ is the Pearson product-moment correlation coefficient.} 
\label{fig:turn_rates}

\end{figure*}

The results of a parameter fit of the first cycle of the above mentioned test flight are shown in Table \ref{tab:turn_rate_law}, where $\rho$ is the Pearson product-moment correlation coefficient between the measured yaw rate and the turn rate estimated by using Eq.~(\ref{eq:psi}) and $\sigma$ is the standard deviation of the estimated turn rate. All data was filtered by calculating a moving average over two seconds before plotting and performing the parameter fitting. 
\begin{table}[h]
\centering
\caption{Fitted turn rate law parameters of the Hydra kite. Values based on the measurements and on the one point and four point kite model.}
\label{tab:turn_rate_law}
\begin{tabular}{ l r r r}
\multicolumn{3}{l}{\textbf{Fitted steering parameters}}\\ 
\hline $ \alphasmax$ [$^o$] & 15.9 & maximal steering~ &~\\ 
$\wrel$ [\%] & 91.0 & \multicolumn{2}{l}{relative kite width ~~4p model}\\ 
$\cs$  & 2.59 & \multicolumn{2}{l}{steering coefficient 1p model} \\ 
$\ctoc$  & 0.93 & \multicolumn{2}{l}{correction factor ~~~~1p model} \\ 
 & ~ \textbf{Measured} & \textbf{1p model} & \textbf{4p model} \\ 
\hline $\ud$ ~[\%] & 26.0 & 26.0 & 26.0 \\ 
$c_0$~~[~\textendash~] & -0.003 & -0.004 & -0.003 \\ 
$c_1$ ~[rad /$m$]& 0.261 & 0.264 & 0.262 \\ 
$c_2$ ~[rad m/$s^2$]& 6.28 & 6.20 & 6.27 \\ 
\hline 
$\rho~~~(PCC)$ & 0.9933 & 0.9999 & 0.9995 \\ 
$\sigma$~~ [rad/s]& 0.002 & 0.0002 & 0.0006 \\ 
\end{tabular} 
\end{table}
The diagrams in Fig.~\ref{fig:turn_rates} illustrate the measured yaw rate, the turn rate estimated by using Eq.~(\ref{eq:psi}) and the relationship between the estimated and measured/ simulated yaw/ turn rates. The term \emph{heading rate} is used for the derivative of the heading angle while the term \emph{yaw rate} is used for the value that was measured by the gyroscope of the inertia measurement unit of the kite that was aligned with its $z$-axis. The numerical derivative of the heading angle of the IMU was too noisy to be used. 
\captionsetup{justification=centering}
\begin{figure*}[htbp]
\centering
\begin{subfigure}{.33\textwidth}
  \centering
  \includegraphics[height = 4.35cm, width=1.08\linewidth]{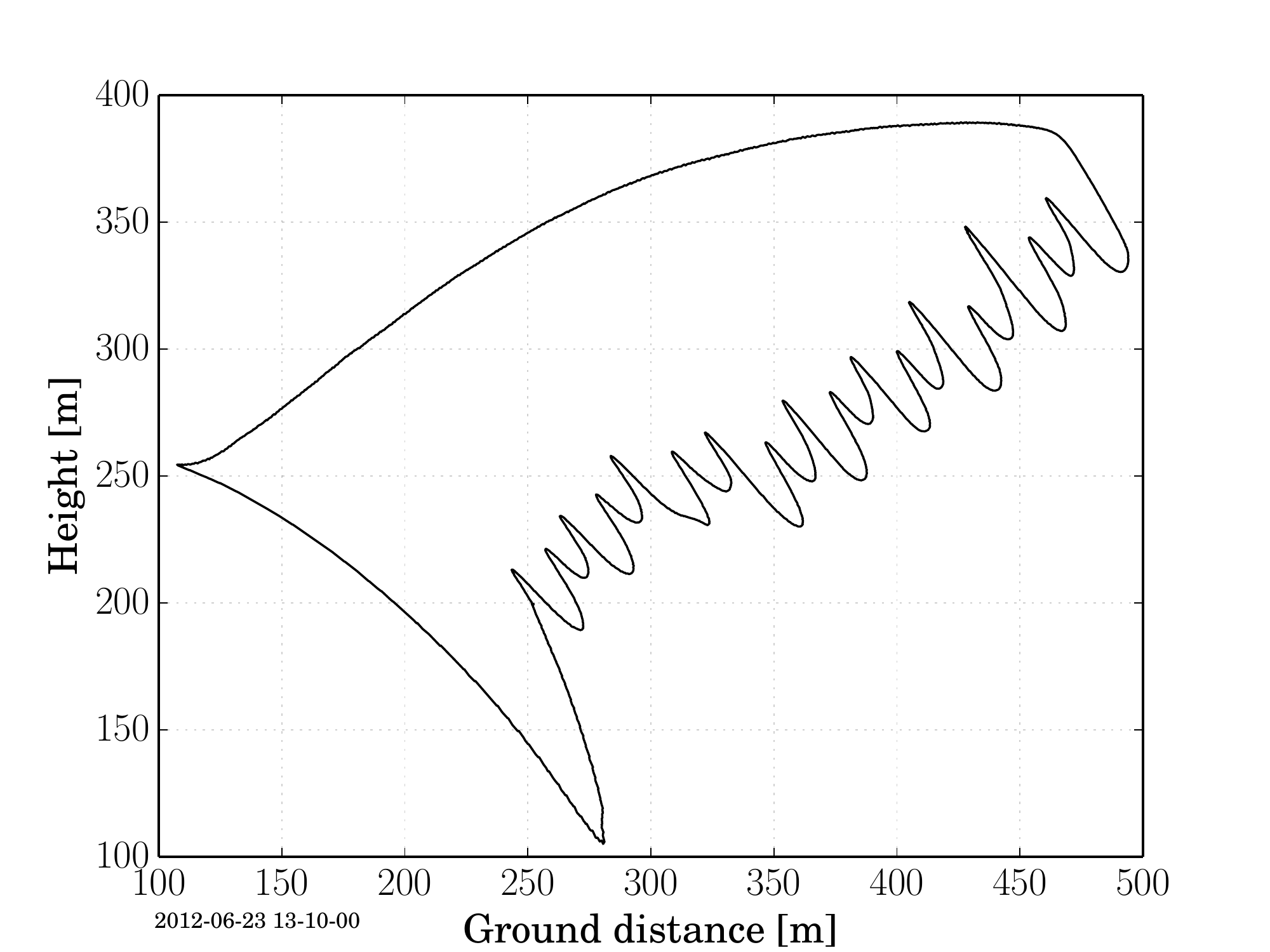}
  \caption{Measured flight path.\\ \mbox{}}
\end{subfigure}
\hfill
\begin{subfigure}{.33\textwidth}
  \centering
  \includegraphics[height = 4.35cm, width=1.08\linewidth]{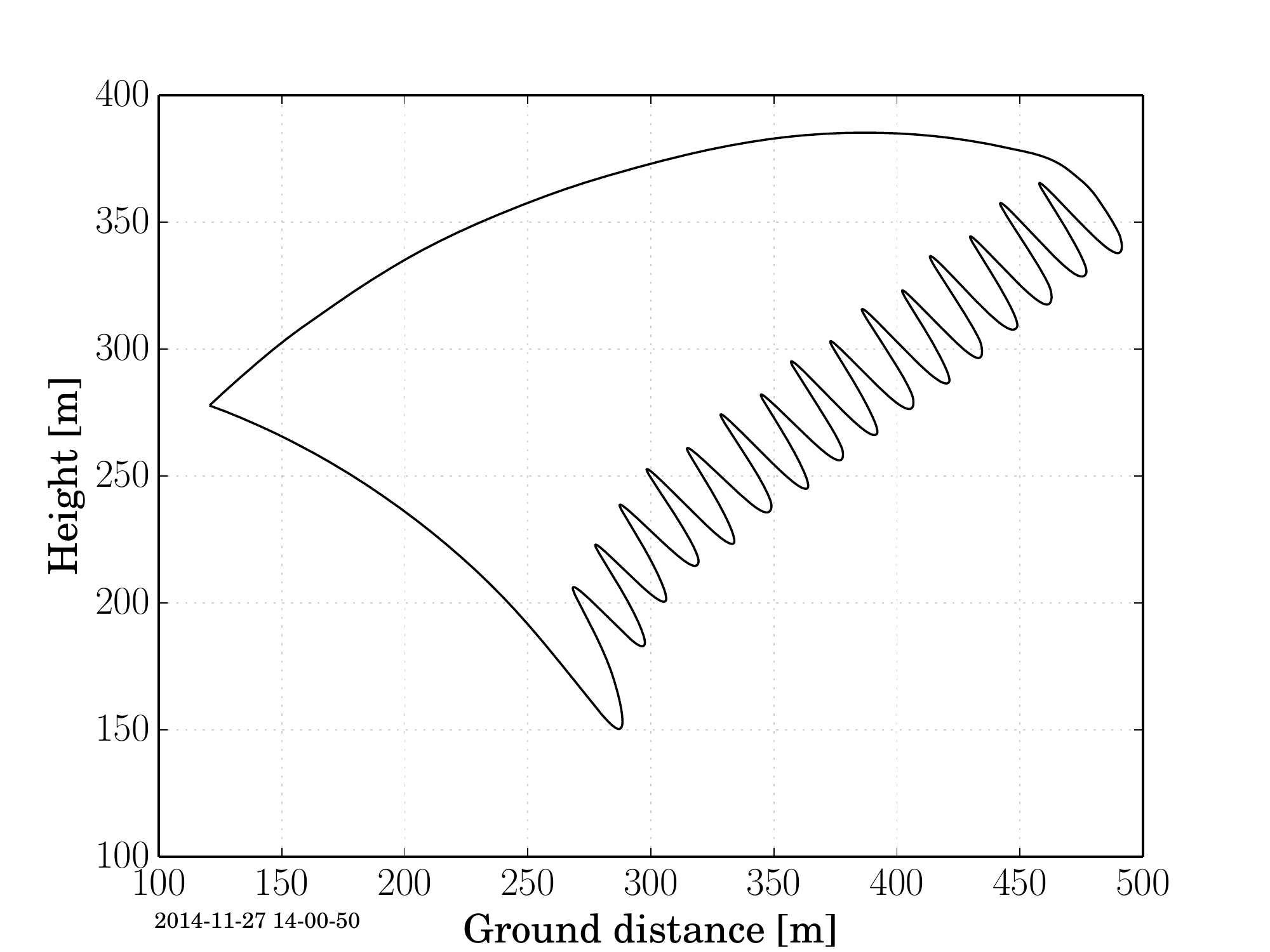}
  \caption{Simulated flight path, four point model. Efficiency error: 1.8\%.}
\end{subfigure}
\begin{subfigure}{.33\textwidth}
  \centering
  \includegraphics[height = 4.35cm, width=1.08\linewidth]{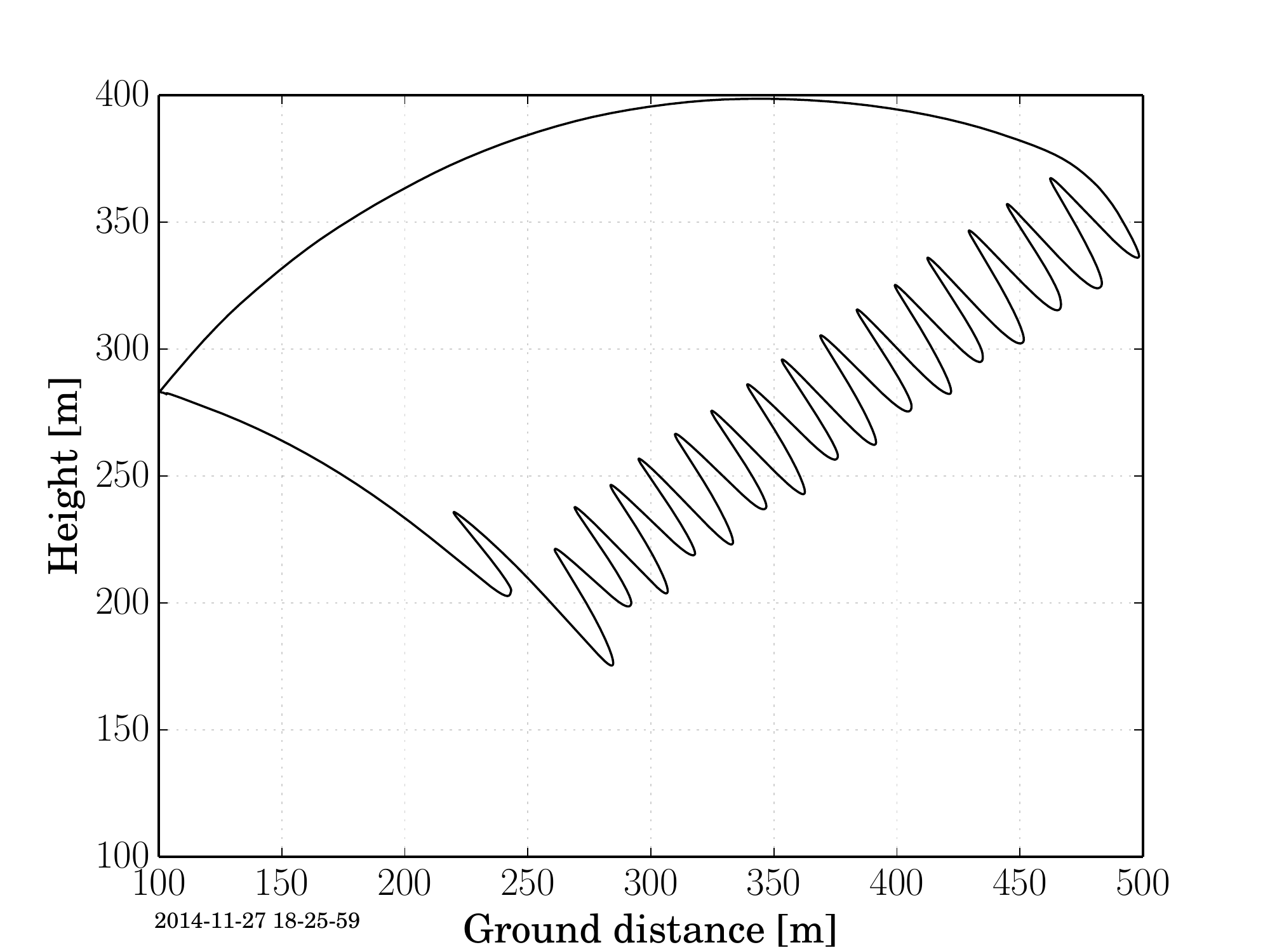}
  \caption{Simulated flight path, point mass model. Efficiency error: 3.1\%. Reel in unstable.}
\end{subfigure}
\caption{Measured and simulated flight paths of one cycle as seen from the side. It can be seen that the minimal and the maximal height are simulated more accurately with the four point model.} 
\label{fig:flight_path}
\end{figure*}
\captionsetup{justification=raggedright}

For the parameter $\kds$, the influence of the depower settings on the steering sensitivity a value of 1.5 was used. This value was estimated based on the geometry of the kite and the bridle. To verify this value, it is suggested to fly figures of eight with a fully depowered kite. This was not done yet in practice. The data measured during the reel-in phase of the kite was not sufficient to validate this parameter because - without flying crosswind - the turn rate of the kite is highly influenced by the turbulence of the wind. The data was too noisy to be useful.

With the point mass model it was difficult to achieve stable parking, using the control parameters of the flight experiment: It was always oscillating around the desired position and therefore flying crosswind even it should not. Therefore the calibration parameters from the four point model had to be used instead.

\subsection{Model comparison}
A first comparison of four model variations (one point kite and four point kite model combined with either a straight or a segmented tether) can be done by parking the kite (steering it towards zenith) in a quasi-steady wind field. 
\begin{table}[h]
\centering
\caption{Comparison of the tether force and the elevation angle of a kite, parking at a line length of 392~m. The simulated results of the one point and the four point model, combined with a straight and a segmented tether are compared.}
\begin{tabular}{  l r r r r}
{{Model}} & {Force [N]} & \textbf{$\sigma_f$} & \textbf{$\beta$ [$ ^{\circ}$] }& \textbf{$\sigma_{\beta}$}\\ 
\hline 
1p, straight tether   & 749.7 &   16.4 &    74.7 &  0.05 \\ 
1p, segmented tether  & 727.5 &    9.2 &    70.7 &  0.02 \\ 
4p, straight tether   & 685.7 &    5.0 &    69.0 &  0.02 \\ 
4p, segmented tether  & 670.2 &    3.2 &    68.5 &  0.02 \\ 
\end{tabular} 
\end{table}
A ground wind speed of 8~m/s and a turbulence intensity of 1\% and an exponential wind profile with  $\alpha = 1/7$ were used for these simulations. 
The difference of the force and of the elevation angle between the most simple and most complex model are about 10\%. Much bigger is the difference in the dynamic behaviour: The variance of the tether force of the four-point model with a segmented tether is more than five times smaller than the variance, using the one point model with a straight tether. The reason for this is the damping, that is induced by the segmented tether and four point kite. 

A realistic model of the non-linear damping of the system is essential for the design of the force controller of the ground station.

\subsection{Results: Power production and flight path}
\noindent When simulating figure of eight flight manoeuvres with the parameters identified in Sect. \ref{sub:paramter_identification} the result as shown in column \emph{Sim. I} in Table~\ref{tab:power_cycles} was disappointing: The computed average power was about 50\% lower than the measured value. To achieve a better match between simulation and measurements it was necessary to increase $\udd$ from 21.3\% to 21.4\% and to decrease the depower setting during reel-in by 2.1\% as shown in column \emph{Sim.~II}. This can be justified first with inaccuracies during the parameter identification and second with a shift of $\udd$ by different apparent wind velocities and/or material creep of the depower/ steering lines. 

The point mass model (Table~\ref{tab:power_cycles} column \emph{Sim. III}) was tuned slightly differently to match the measured power output and to achieve a similar flight trajectory. Nevertheless, the errors between the one point kite model and the measurements were higher, for example an error of 3.1\% instead of 1.8\% for the cycle efficiency $\etacyc$.
\begin{table}[h]
\centering
\caption{Parameters of measured and simulated pumping cycles. The lowest efficiency error is achieved with the four point kite simulation Sim. II. The cycle efficiency $\etacyc$ is the product of the pumping efficiency $\etap$ and the duty cycle~\cite{Fechner2013}.}
\label{tab:power_cycles}
\begin{tabular}{  l r r r r}
\multicolumn{2}{r}{~~\bfseries{Measured}} & \textbf{Sim. I} & \textbf{Sim. II} & \textbf{Sim. III}\\ 
\hline 
$\vwg$ [m/s]     & 9.51    &    9.51 &    9.51 &  9.51 \\ 
$u_\mr{d,ri} [\%] $  & 42.2~~     &   42.2~~   &    40.1~~ &  44.1~~\\ 
$\udd$ [\%]   & -~~     &   21.30 &   23.40 & 20.80\\ 
L/D, reel-out    & -~~     &    4.13 &    4.64 &  4.53\\ 
$\fto$ [N]    & 2942.~~~& 2213.~~~& 2876.~~~& 2956.~~~\\ 
$\fti$ [N]    &  653.~~~&  379.~~~&  600.~~~& 570.~~~\\ 
$\vto$ [m/s]  & 1.99    &    1.20 &    1.89 &   1.88\\ 
$\vti$ [m/s]  & -7.28   &   -7.22 &   -7.69 &  -7.66\\ 
$\pav$  [W]    & 3726.40 & 1953.10 & 3681.50 & 3735.80\\ 
$\etap$ [\%]  & 79.10   &   83.00 &   79.70 &   81.70 \\ 
duty cycle [\%]  & 78.70   &   85.30 &   80.30 &   80.40\\
$\etacyc$ [\%]& 62.20   &   70.80 &   64.00 &   65.70\\ 
\end{tabular} 
\end{table}
(The value $\pav$ is the average mechanical power over the whole cycle, and $\etacyc$ is the cycle efficiency, the quotient of the mean mechanical power and the average mechanical reel-out power).

A two dimensional projection of flight trajectory, height of the kite vs. the ground distance, is a suitable means for visualisation and comparison of different flights. In Fig. \ref{fig:flight_path} the measured and the simulated flight path of one cycle is shown. The maximum height differs by less then 5\%. The minimum height differs by about 45~m. One reason for this are the inaccuracies of the Global Navigation Satellite System (GNSS) based height measurement. 
%

\section{Conclusions}
\label{sec:summary_and_conclusion}
\noindent The computed dynamic response of the kite to steering inputs compares well to the response measured during test flights. In all situations the estimated turn rate of the wing was within $\pm 14\%$ of the full range of the experimentally measured values while the standard deviation was only $\pm 0.1\%$ of the full range.
Similar results, but limited to the steering of a kite on a tether of a fixed length were presented in \cite{Erhard2013} and \cite{Fagiano2013}.

By modifying empirically the parameters $\cs$ and $\ctoc$, the proposed point mass model can be adapted to match all parameters of the turn rate law. Compared to the proposed four point model it runs faster but is less accurate and can become dynamically unstable at low tether forces. 

In \cite{Jehle2014} it was assumed that the turn-rate law derived in \cite{Erhard2013} and \cite{Fagiano2013} would only be valid for ram-air kites. We found, that it is valid for Leading Edge Inflatable tube kites, too.

The parameters of the mechanistic four point model can be derived from the physical properties of any soft kite and any asynchronous generator. Only small changes are required for other kites and generators. It is well suited for controller development and can be used not only for the pumping cycle operation of the kite, but also for the simulation of launching, landing and airborne parking.

For a full model validation of a specific system, two enhancements of the test design are needed: First, accurate wind measurements at the height of the kite. Second, an accurate measurement of the maximum and minimum L/D of the kite and of the depower offset $\udd$. 

\sloppy The presented models have shown to be easily adaptable and well suited for flight path optimization and the development of KPS estimators and KPS controllers. 
While the corrected one-point model with an adapted flight path controller can be sufficient for flight-path optimization, the four point model is better suited for controller validation in a broader range of flight conditions.

Even though the accuracy of the predicted power output is not yet sufficiently validated, the one point model, using the correction according to Eq. \ref{eq:correction} is predicting the influence of gravity on the turn rate much better than uncorrected point mass models and the four point model has a much more realistic dynamic response to the steering input then simpler models while still being real-time capable. The source code is published under the GNU LGPL License in the context of the FreeKiteSim~\cite{Fechner2014} project.

\section*{Acknowledgements}  
\noindent {\small The authors want to thank Filip Saad and Rachel Leudhold for their participation in the development and documentation of a sport kite simulator, which in turn inspired the development of the KPS simulator as described in this paper. In addition they want to thank William Anderson and Axelle Vir\'e for proof reading.}

\bibliographystyle{elsarticle-num}







\end{document}